\documentclass[pdftex,twocolumn,epjc3]{svjour3}         
\pdfoutput=1
\journalname{Eur. Phys. J. C}
\usepackage{color}
\usepackage{graphicx}
\usepackage{wrapfig,enumerate,slashed}
\usepackage{paralist}
\usepackage{multirow}
\usepackage{footmisc}
\usepackage{amsmath,amssymb}
\usepackage{wasysym}
\usepackage{graphicx}
\usepackage{color}
\usepackage[alsoload=hep]{siunitx}

\RequirePackage[numbers,sort&compress]{natbib}
\RequirePackage[colorlinks,citecolor=blue,urlcolor=blue,linkcolor=blue]{hyperref}

\newcommand{\be}{\begin{eqnarray}}
\newcommand{\ee}{\end{eqnarray}}

\newcommand{\bee}{\begin{eqnarray}}
\newcommand{\eee}{\end{eqnarray}}
\newcommand{\beeq}{\begin{equation}}
\newcommand{\eeeq}{\end{equation}}

\newcommand{\fb}{{\text{fb}}}

\renewcommand{\vec}{\bf}

\def\etac{\ensuremath{\raise.4ex\hbox{$\eta$}_{{c}}}}

\def\Jpsi{\ensuremath{J/\psi}}

\begin{document}

\title{Measuring rare and exclusive Higgs boson decays into light resonances}

\author{Andrew S. Chisholm\thanksref{e1,addr1}
        \and
        Silvan Kuttimalai\thanksref{e2,addr2}
        \and
        Konstantinos Nikolopoulos\thanksref{e3,addr1}
        \and
        Michael Spannowsky\thanksref{e4,addr2}
}
\thankstext{e1}{Andrew.Chisholm@cern.ch}        
\thankstext{e2}{s.s.kuttimalai@durham.ac.uk}    
\thankstext{e3}{k.nikolopoulos@bham.ac.uk}      
\thankstext{e4}{michael.spannowsky@durham.ac.uk}

\institute{School of Physics and Astronomy,\\University of Birmingham, B15 2TT, United Kingdom\label{addr1}
          \and
          Institute for Particle Physics Phenomenology, Department
          of Physics,\\Durham University, DH1 3LE, United Kingdom\label{addr2}
}


\maketitle

\begin{abstract}
We evaluate the LHC's potential of observing Higgs boson decays into light elementary or
composite resonances through their hadronic decay channels. We focus on the Higgs boson production processes with the largest cross sections, $pp \to h$
and $pp \to h+\mathrm{jet}$, with subsequent decays $h \to ZA$ or $h
\to Z\,\eta_c$, and comment on the production process $pp \to hZ$. By exploiting track-based jet substructure observables and extrapolating to $3000~\mathrm{fb}^{-1}$
we find ${\cal BR}(h \to ZA) \simeq {\cal BR}(h \to Z \eta_c) \lesssim 0.02$ at 95~\% CL. We interpret this limit in terms of the 2HDM Type 1. We find that
searches for $h\to ZA$ are complementary to existing measurements and can constrain large parts of the currently allowed parameter space.
\end{abstract}

\section{Introduction}
\label{sec:intro}
The greatly successful Run 1 of the large hadron collider (LHC) culminated in the discovery of a
state that resembles the standard model (SM) Higgs boson
\cite{Aad:2012tfa,Chatrchyan:2012xdj}. First measurements of its
couplings to gauge bosons and third-generation fermions are in good
agreement with SM predictions \cite{Khachatryan:2016vau}.
However, the current precision of the measurement of Higgs boson couplings
and properties cannot rule out Higgs boson decays into light
resonances. In the SM, examples of such light resonances include the composite
unflavoured mesons and quarkonium states, e.g. the $\Jpsi$.

Furthermore, Higgs boson decays into elementary light resonances are
predicted by many extensions of the SM~\cite{Curtin:2013fra}. They arise generically in scenarios with
multiple Higgs fields or kinetic mixing between SM gauge bosons and
bosons of a dark $U(1)$ gauge group. In the NMSSM, Higgs boson decays into an additional
light CP-odd scalar can occur. Close to the alignment limit of the
Two-Higgs-Doublet Model (2HDM) of Type I or II, a light CP-odd scalar with
mass of few \si{\giga\electronvolt} can also be phenomenologically accommodated with a \SI{125}{\giga\electronvolt}
SM-like Higgs boson $h$ \cite{Bernon:2015qea}. Higgs boson decays
into vector bosons of the SM and an additional
spontaneously broken $U(1)_D$ \cite{Curtin:2014cca} can arise
through 
kinetic mixing induced by heavy particles that carry
hypercharge, e.g. $h \to Z Z_D$ or $h \to \gamma Z_D$.

Searches for light composite resonances have been proposed to set a
limit on the Higgs boson couplings to first and second-generation
quarks \cite{Bodwin:2013gca,Kagan:2014ila}. However, for SM couplings the branching ratios for exclusive Higgs boson decays
are generally of $\mathcal{O}(10^{-5})$ or less
\cite{Bodwin:2013gca,Isidori:2013cla,Koenig:2015pha}, e.g.
$\mathcal{BR}(h \to Z\,\eta_c) \simeq 1.4 \times 10^{-5}$,
$\mathcal{BR}(h \to \rho^0 \gamma) \simeq 1.68 \times 10^{-5}$ or
$\mathcal{BR}(h \to J/\psi~\gamma) \simeq 2.95 \times 10^{-6}$,
resulting in small expected event yields. Nevertheless, both general purpose experiments at the LHC
 have performed searches for exclusive Higgs
boson decays, focusing on the dimuon decays of vector quarkonia. With Run 1
data the ATLAS collaboration has set 95~\% confidence level (CL) upper limits of $\mathcal{O}(10^{-3})$ on the branching ratios for $\mathcal{BR}(h \to J/\psi~\gamma)$ and
$\mathcal{BR}(h \to \Upsilon(\mathrm{1S,2S,3S})~\gamma)$~\cite{Aad:2015sda}, while the CMS collaboration obtained a similar upper limit for $\mathcal{BR}(h \to J/\psi~\gamma)$~\cite{Khachatryan:2015lga}. Recently, the ATLAS collaboration has also set a 95~\% CL upper limit of $1.4 \times 10^{-3}$ on $\mathcal{BR}(h \to \phi~\gamma)$~\cite{ATLASphi}.

Hence, rare decays of Higgs bosons into light elementary or composite
resonances are of direct relevance for the two most important tasks of
the upcoming LHC runs:
\begin{inparaenum}[(a)]
\item precision measurements of the Higgs boson properties; and
\item searches for new physics.
\end{inparaenum}

While most existing search strategies rely upon resonance decays
into leptons, i.e. muons, the total width of most composite resonances and elementary
scalars is dominated by decays into hadronic final states, e.g.
$\mathcal{BR}(\etac \to \mathrm{hadrons}) > 52~\%$\footnote{Based on a simple sum of the branching fractions for the observed decays of the $\etac$ into stable hadrons.}~\cite{Agashe:2014kda}.
Instead of exploiting only leptonic decay modes, we therefore propose that the inclusive
hadronic decays be considered. Light resonances $X$ with masses of
$m_X=1-\SI{10}{\giga\electronvolt}$ produced in decays of the Higgs boson with
a mass of \SI{125}{\giga\electronvolt}, are highly boosted and their decay products
are thus confined within a small area of the detector. The angular
separation of the decay products of the resonance $X$ scales like
$\Delta R=\sqrt{\Delta\eta^{2}+\Delta\phi^{2}} \sim 4 m_X / m_h$,
where $\eta$ is the pseudorapidity and $\phi$ the azimuthal angle.
Separating the decay products in the
calorimeters of the detector poses a challenge, as the typical size
of hadronic calorimeter cells is $0.1\times 0.1$ in the $(\eta,\phi)$ plane.

Thus, to discriminate two jets the angular separation of their axes
has to be roughly $\Delta R \gtrsim 0.2$. If opening angles are
smaller, the total energy deposit of the resonance decay products
can still be measured, but the substructure, i.e. the energy sharing
between the decay products, becomes opaque. To maintain the ability to
separate between signal and QCD-induced backgrounds we propose to
utilise track-based reconstruction. Trajectories of charged
particles as measured in the tracking detectors provide a much better
spatial resolution than the reconstructed calorimeter clusters.
Recently, a similar approach was advocated for highly boosted
electroweak scale resonances \cite{Katz:2010mr,Schaetzel:2013vka,Larkoski:2015yqa,Spannowsky:2015eba}, for which dedicated taggers
have been developed.\footnote{\url{http://www.ippp.dur.ac.uk/~mspannow/webippp/HPTTaggers.html}}

In this work, we use track-based reconstruction techniques to
evaluate the sensitivity of general purpose detectors at hadron colliders, with
characteristics similar to those of ATLAS~\cite{Aad:2008zzm} and
CMS~\cite{Chatrchyan:2008aa}, in measuring rare Higgs boson decays
into light hadronically decaying resonances. Focusing on the High
Luminosity LHC (HL-LHC) regime, our analysis assumes a dataset
corresponding to an integrated luminosity of \SI{3000}{\per\fb}
collected at center-of-mass energy $\sqrt{s}=\SI{13}{\tera\electronvolt}$.
We consider two production channels for the Higgs boson:
inclusive Higgs boson production and Higgs boson production in association with a
hard jet of transverse momentum $p_{T}>\SI{150}{\giga\electronvolt}$.

As two benchmark
cases for rare Higgs boson decays into light resonances we consider $h
\to Z(\to\ell\ell)+\etac$ and $h \to Z(\to\ell\ell)+A$, where $A$ is assumed to
be an elementary CP-odd scalar of mass \SI{4}{\giga\electronvolt}
which decays mostly hadronically. The presence of two high-pT isolated leptons from the Z boson decay, ensure an efficient trigger strategy for HL-LHC environment. The characteristics of the $h\to Z(\to\ell\ell)+\etac$ benchmark
are expected to be representative of similar decays to vector charmonia (e.g. $h\to Z(\to\ell\ell)+\Jpsi$),
due to similarities in their hadronic decay patterns and small mass differences relative to the scale
of the jet momenta relevant in the decays of Higgs boson with a mass of \SI{125}{\giga\electronvolt}.

The event generation is described in Sect.~\ref{sec:eventgen}, while
Sect.~\ref{sec:sel} is devoted to the details of the reconstruction of
the Higgs boson decay products and event selection. The statistical
analysis and expected sensitivity are given in Sect.~\ref{sec:results}.
In Sect.~\ref{sec:int} the expected results are interpreted
in terms of 2HDM models. We offer a summary of our findings in
Sect.~\ref{sec:sum}.

\section{Event generation}
\label{sec:eventgen}

For the simulation of both the signal and the background contributions
we employ a modified version of Sherpa 2.2.0 \cite{Gleisberg:2008ta}
that was adapted in such a way as to facilitate the
simulation of Higgs decays into composite resonances. Parton
shower effects, hadronisation, as well as underlying event
contributions are taken into account throughout. Both Higgs boson
production processes, $h+\rm{jet}$ and inclusive $h$, are calculated
at NLO and matched to the parton shower. Finite top quark mass effects
in the gluon fusion production mechanism are taken into account as
described in Ref.~\cite{Buschmann:2014sia}. The Higgs boson decays
$h\rightarrow Z\,\etac$, $h\rightarrow ZA$ as well as the subsequent
decay of the pseudoscalar and the $Z$ boson are calculated perturbatively at
leading order using the algorithm and methods described in Ref.~\cite{Hoche:2014kca}. Spin-correlations are thus retained in all
resonance decays. The UFO model format, supported by Sherpa, was used
for the implementation of an elementary pseudoscalar and its interactions
\cite{Hoche:2014kca,Degrande:2011ua}.

The $Z+\rm{jets}$ production is expected to represent the dominant
background in this search with other contributions such as $t\bar{t}$
production being suppressed to a negligible level by requiring an
opposite-charge same-flavour dilepton with an invariant mass
consistent with the $Z$ boson mass.  For inclusive $Z$ boson
production ($Z+\rm{jets}$), we take into account the full dilepton
final state in the matrix elements and calculate the core process at
NLO. We account for additional hard jet emissions by means of multijet
merging techniques \cite{Hoeche:2012yf} and include leading order
matrix elements with up to two additional jets in the setup.

We process the generated events with the DELPHES fast simulation
framework~\cite{deFavereau:2013fsa}, which uses parametrised
descriptions of the response of particle physics detectors to provide
reconstructed physics objects, allowing a realistic data analyses to be performed. As
an example of a general purpose LHC detector, the default ATLAS
configuration card included in DELPHES is used.

\section{Reconstruction setup and selection}
\label{sec:sel}

\subsection{Leptonic $Z$ boson decay reconstruction}
\label{sec:zreco}
The reconstruction of $Z\to\ell\ell$ decays begins with the identification of isolated lepton (electron or muon) candidates. Reconstructed
leptons are required to satisfy $p_{T} > \SI{8}{\giga\electronvolt}$ and $|\eta|<2.5$, one lepton is required to fulfill a trigger requirement of $p_{T} > \SI{25}{\giga\electronvolt}$. An isolation requirement based on the presence of reconstructed tracks
and calorimeter deposits within $\Delta R < 0.2$ of a lepton is imposed. The sum of the transverse momentum of such objects is required to be less
than $10~\%$ of the $p_T$
of the lepton itself. Oppositely charged pairs of isolated leptons, which satisfy $\SI{81}{\giga\electronvolt} < m_{\ell\ell} < \SI{101}{\giga\electronvolt}$
are identified as $Z$ boson candidates.

\subsection{Hadronic resonance reconstruction}
\label{sec:resoreco}
The reconstruction of hadronically decaying resonances within events begins with a search for anti-$k_{t}$ calorimeter jets with $R=0.4$,
seeded by clusters of calorimeter energy deposits. Calorimeter jets are required to have $p_{T} > \SI{30}{\giga\electronvolt}$ and $|\eta|<2.5$. Any jets which
are within $\Delta R < 0.3$
of leptons forming a $Z\to\ell\ell$ candidate are rejected. Following the identification of such a jet, the jet constituents are used to seed a search
for an anti-$k_{t}$ calorimeter jet with $R=0.2$. The identification of an $R=0.2$ jet from the constituents of the initial $R=0.4$ jet is
required to be successful. This procedure, i.e. the reconstruction of anti-$k_{t}$ $R=0.2$ jets from the constituents of identified $R=0.4$ jets,
is repeated for track jets, seeded by reconstructed charged particles. Track jets are associated to calorimeter jets by a simple spatial matching,
based on a requirement of $\Delta R < 0.4$ between the axes of the $R=0.4$ calorimeter and track jets.
Only jets reconstructed with both calorimeter and track components are considered for further analysis and at least one such jet is required to be reconstructed.

To distinguish hadronically decaying charmonium states or light scalars from the copious production of low $p_{T}$ jets, a boosted decision tree (BDT) is used through the TMVA package~\cite{Hocker:2007ht}.
The following variables are used as input to the BDT:
\begin{itemize}
\item The $p_T$ of the $R=0.4$ track and calorimeter jets, as the Higgs boson decay products are expected to have a harder jet pT spectrum.
\item The masses of the $R=0.4$ and $R=0.2$ track and calorimeter jets, as the jets in the signal are expected to be close to the mass of the light resonance.
\item The number of track constituents associated with the $R=0.4$ and $R=0.2$ track jets, as the signal is expected to have a lower track multiplicity given the upper bound imposed by the light resonance mass.
\item The ratio of the $R=0.2$ calorimeter (track) jets $p_T$ to the $p_T$ of the associated $R=0.4$ calorimeter (track) jet, this quantity is expected to prefer values more toward unity in the signal case where a narrow boosted topology is expected, a wider  distribution expected from the QCD jet background.
\item The spatial separation, $\Delta R$, between the leading $p_{T}$ track within the $R=0.4$ track jet and the jet axis.
\item The ratio of the highest track $p_{T}$ to the $p_{T}$ of the $R=0.4$ track jet.
\end{itemize}
The final variables are designed to exploit the fact that in the signal we find on average fewer charged tracks and, due to the very small resonance mass, a smaller active area of the jet.

\begin{figure*}[h]
\centering
\includegraphics[width=0.40\textwidth]{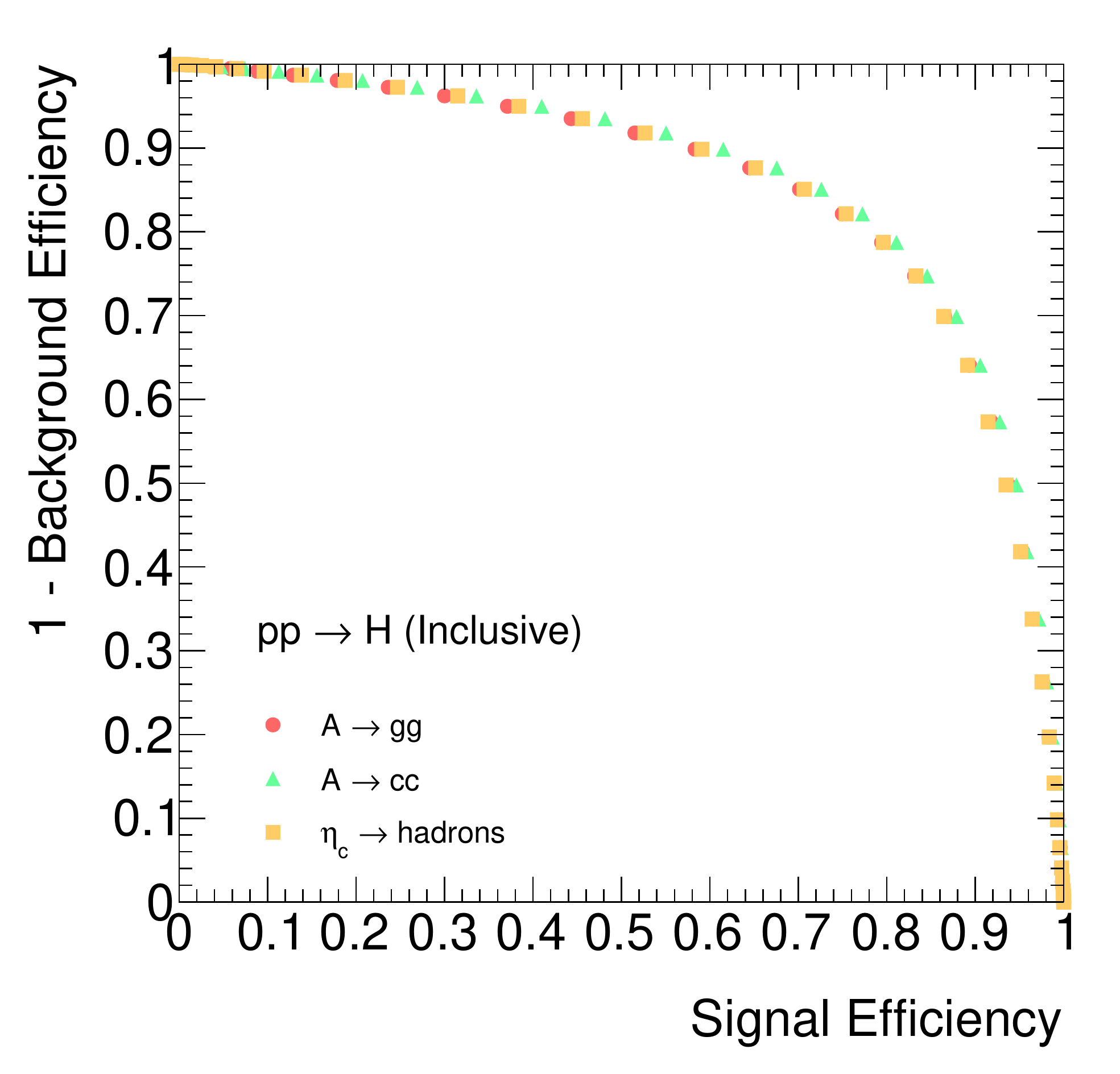}
\includegraphics[width=0.40\textwidth]{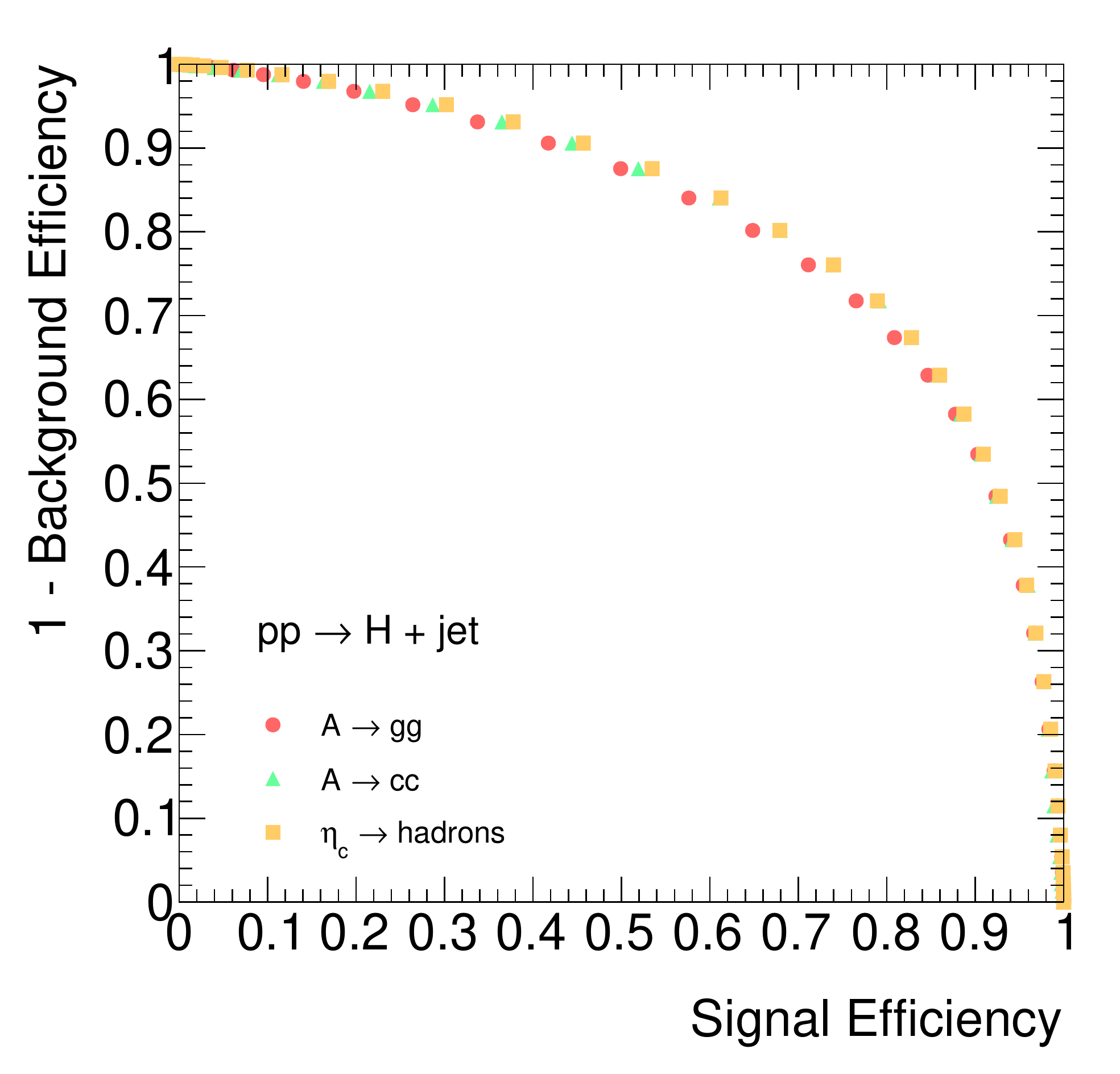}
\caption{The background rejection as a function of signal efficiency for the low mass resonances
considered for the inclusive (left) and $h+\rm{jet}$ (right) production channels.\label{fig:roc}}
\end{figure*}

\begin{figure*}[h]
\centering
\includegraphics[width=0.24\textwidth]{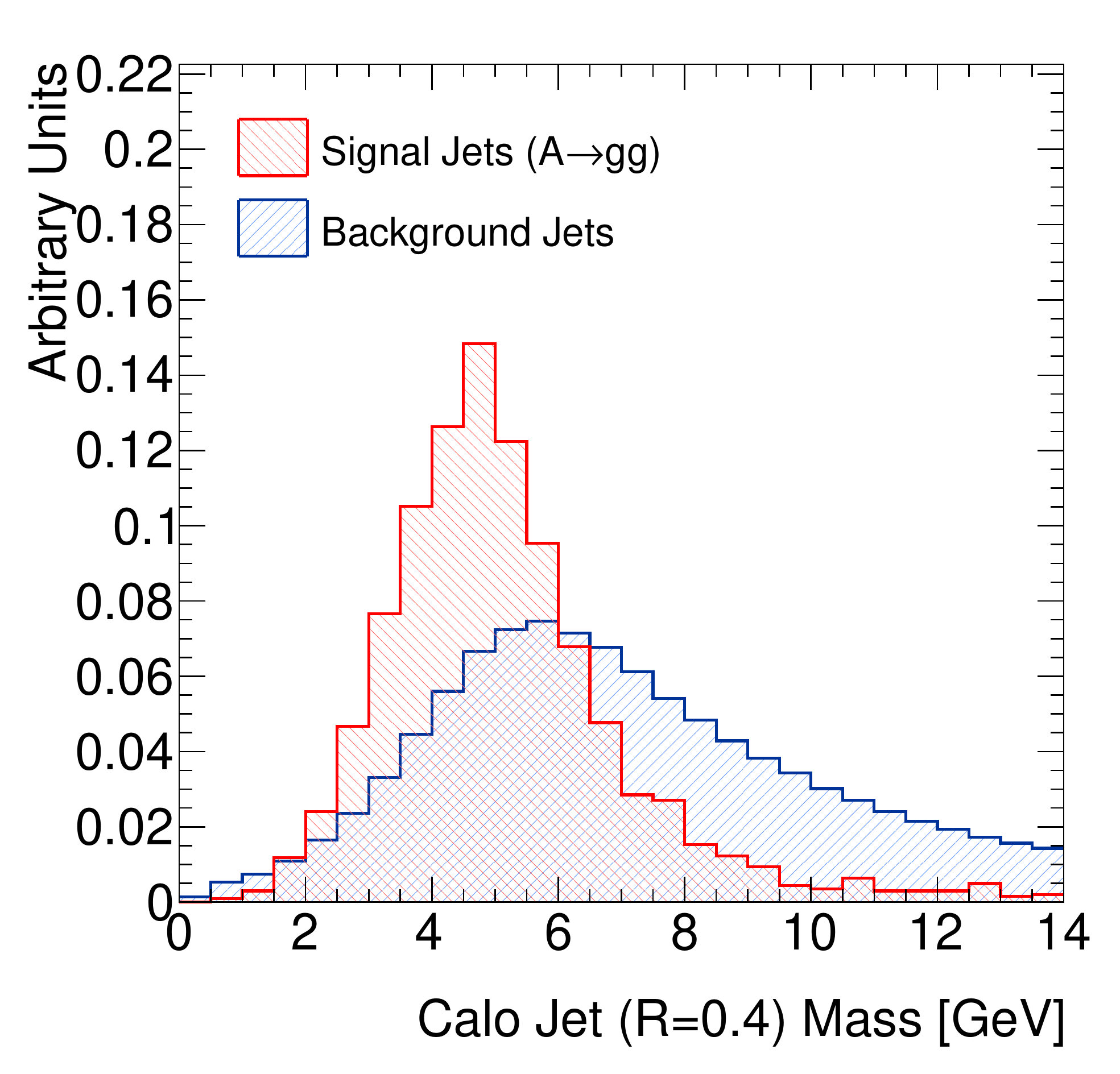}
\includegraphics[width=0.24\textwidth]{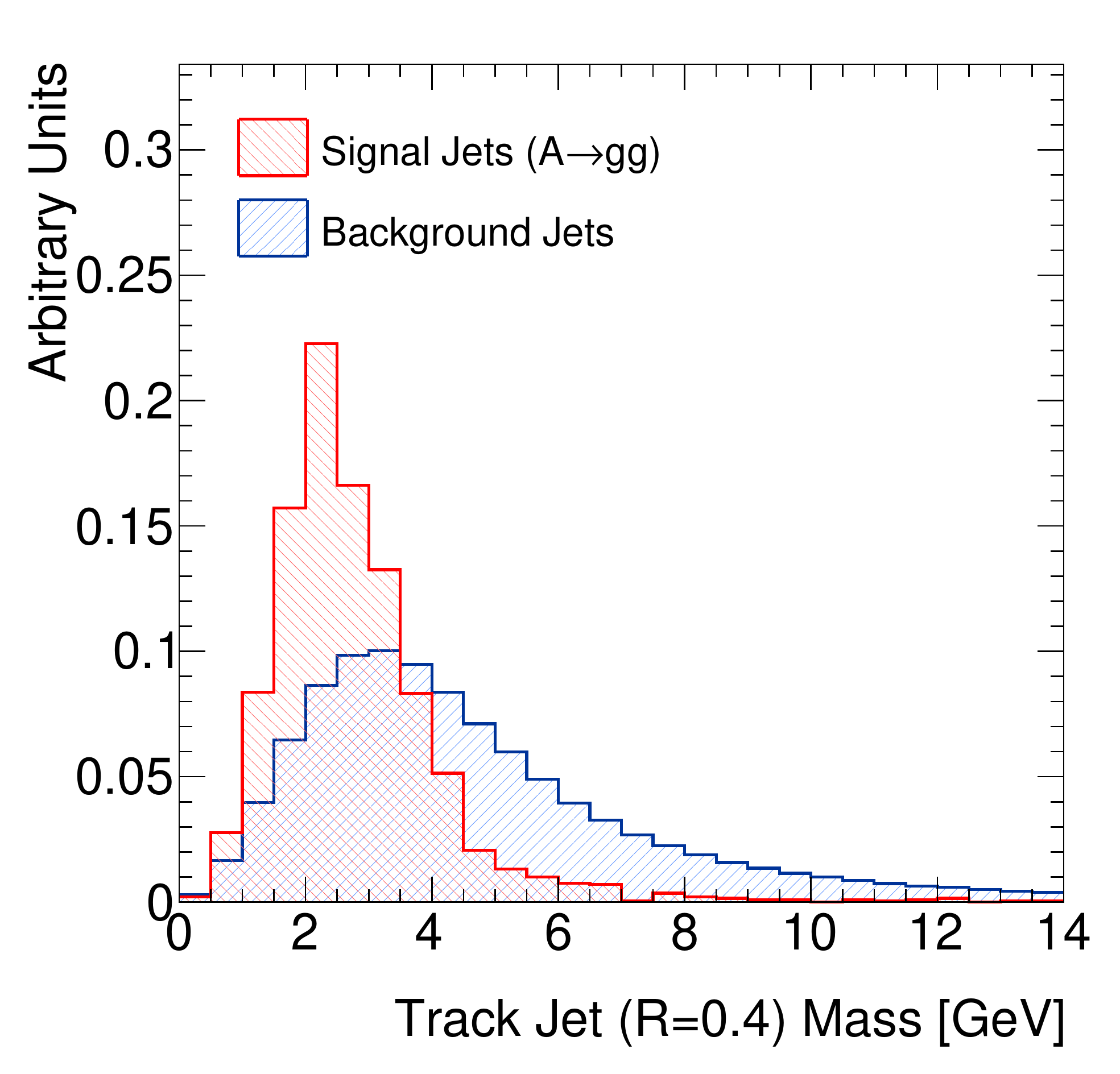}
\includegraphics[width=0.24\textwidth]{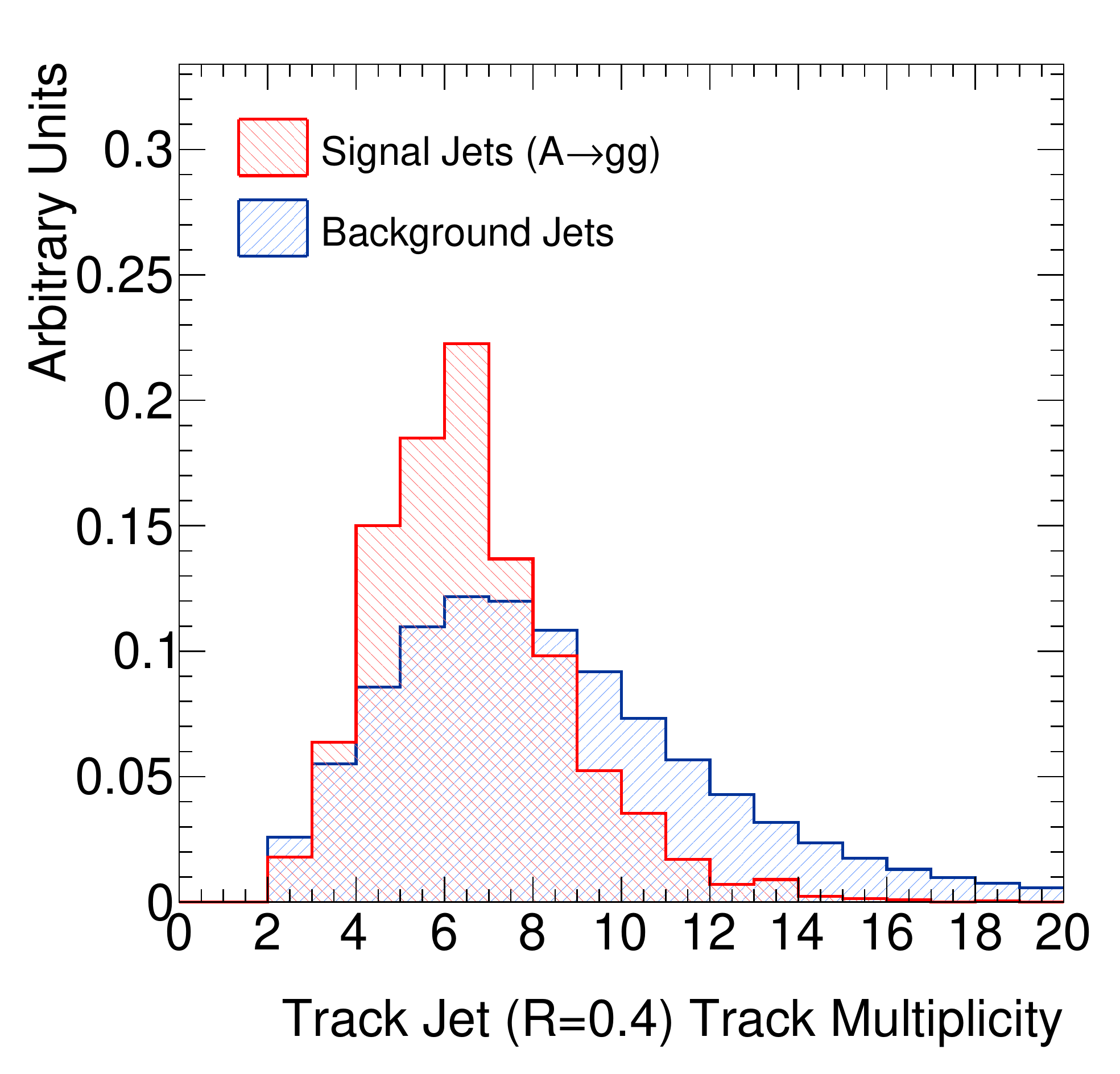}
\includegraphics[width=0.24\textwidth]{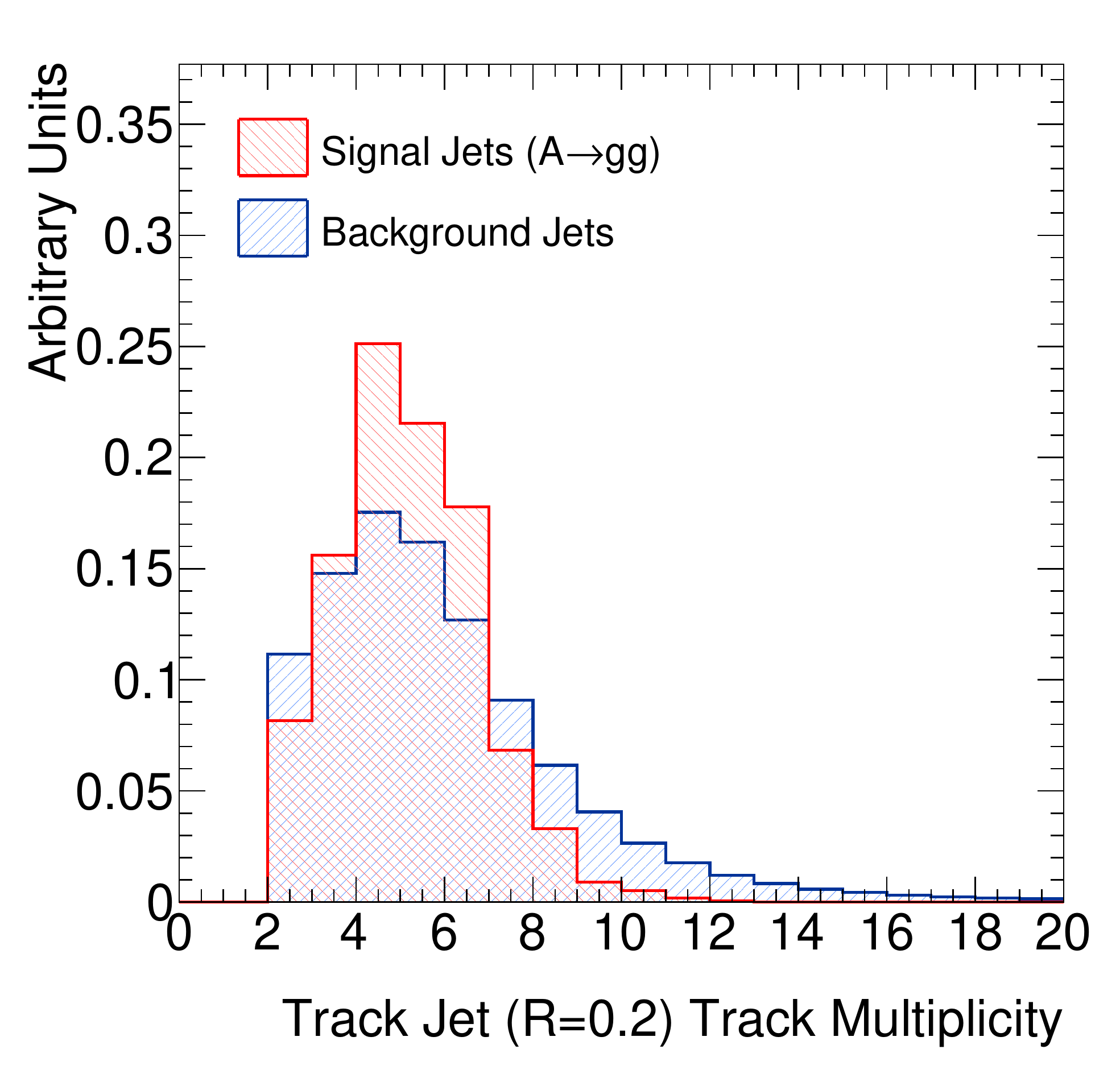}
\includegraphics[width=0.24\textwidth]{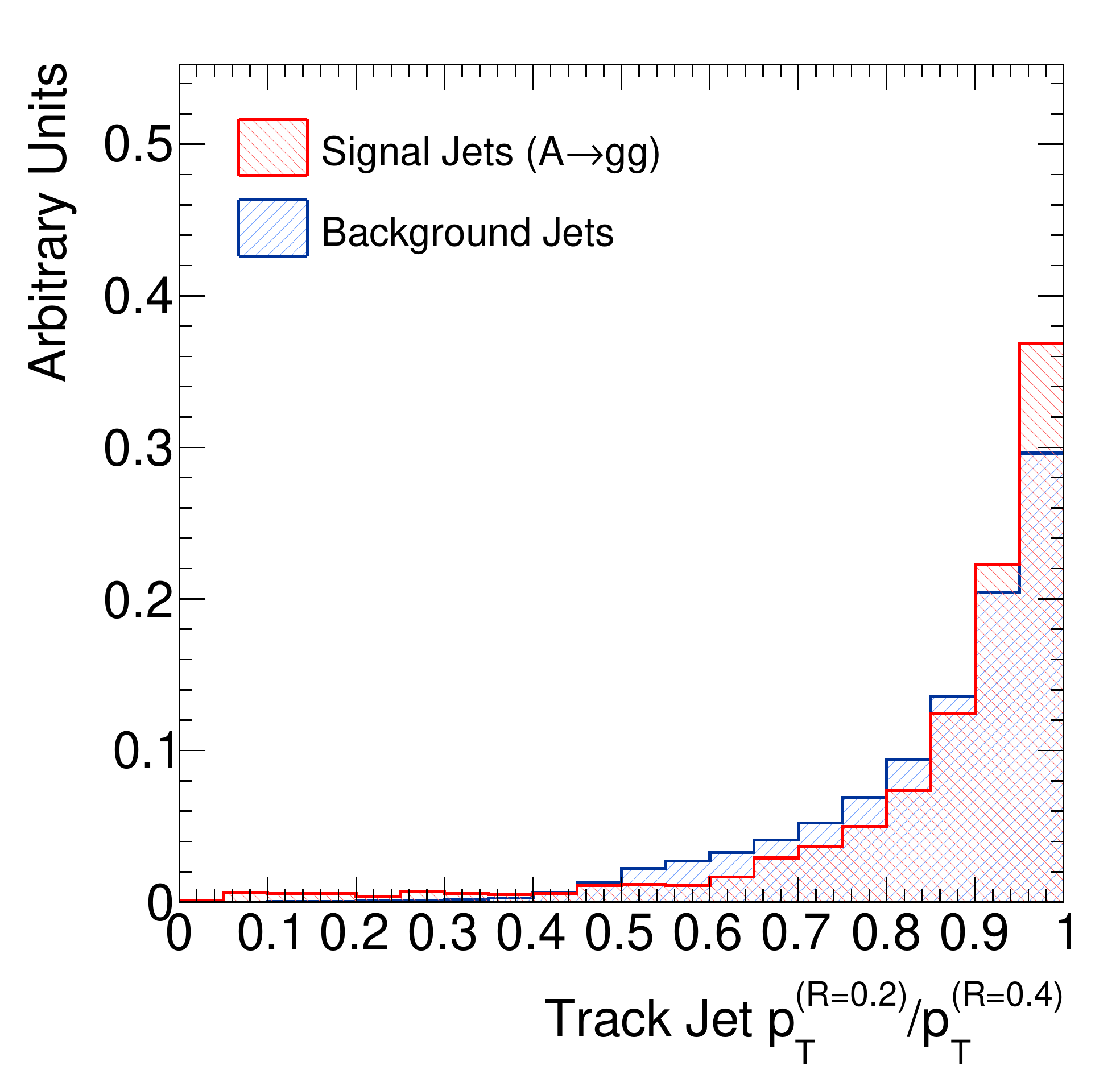}
\includegraphics[width=0.24\textwidth]{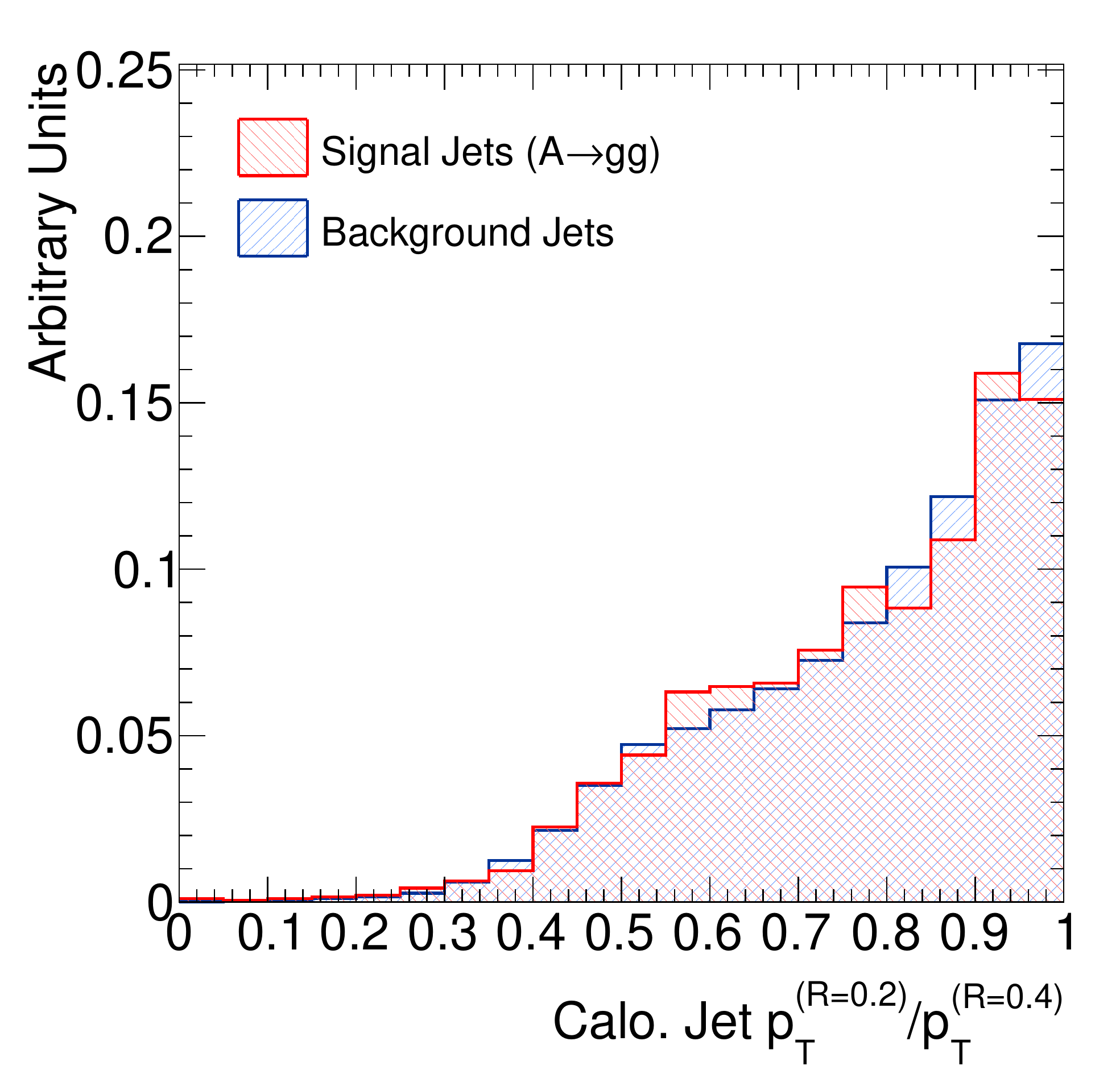}
\includegraphics[width=0.24\textwidth]{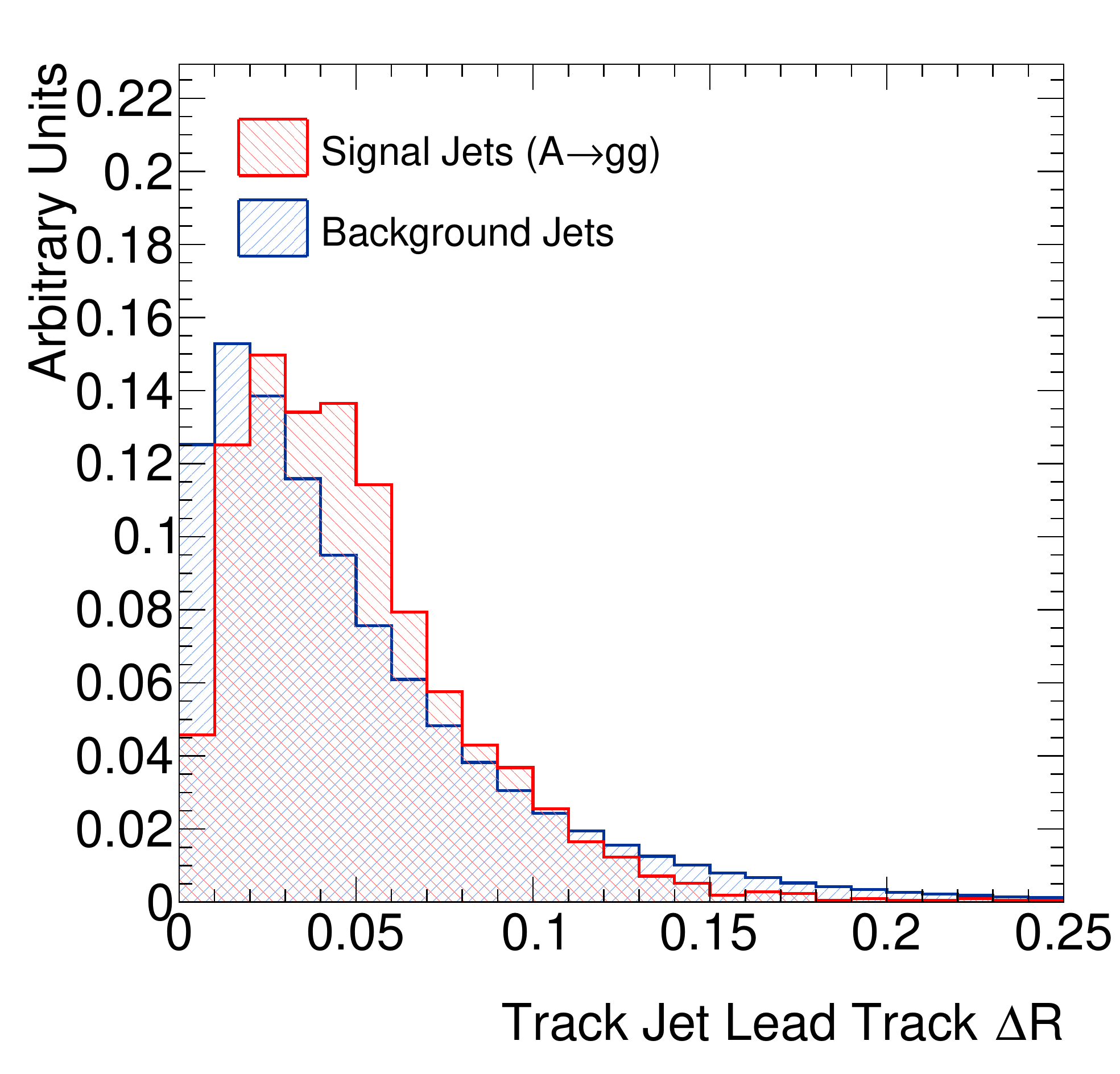}
\includegraphics[width=0.24\textwidth]{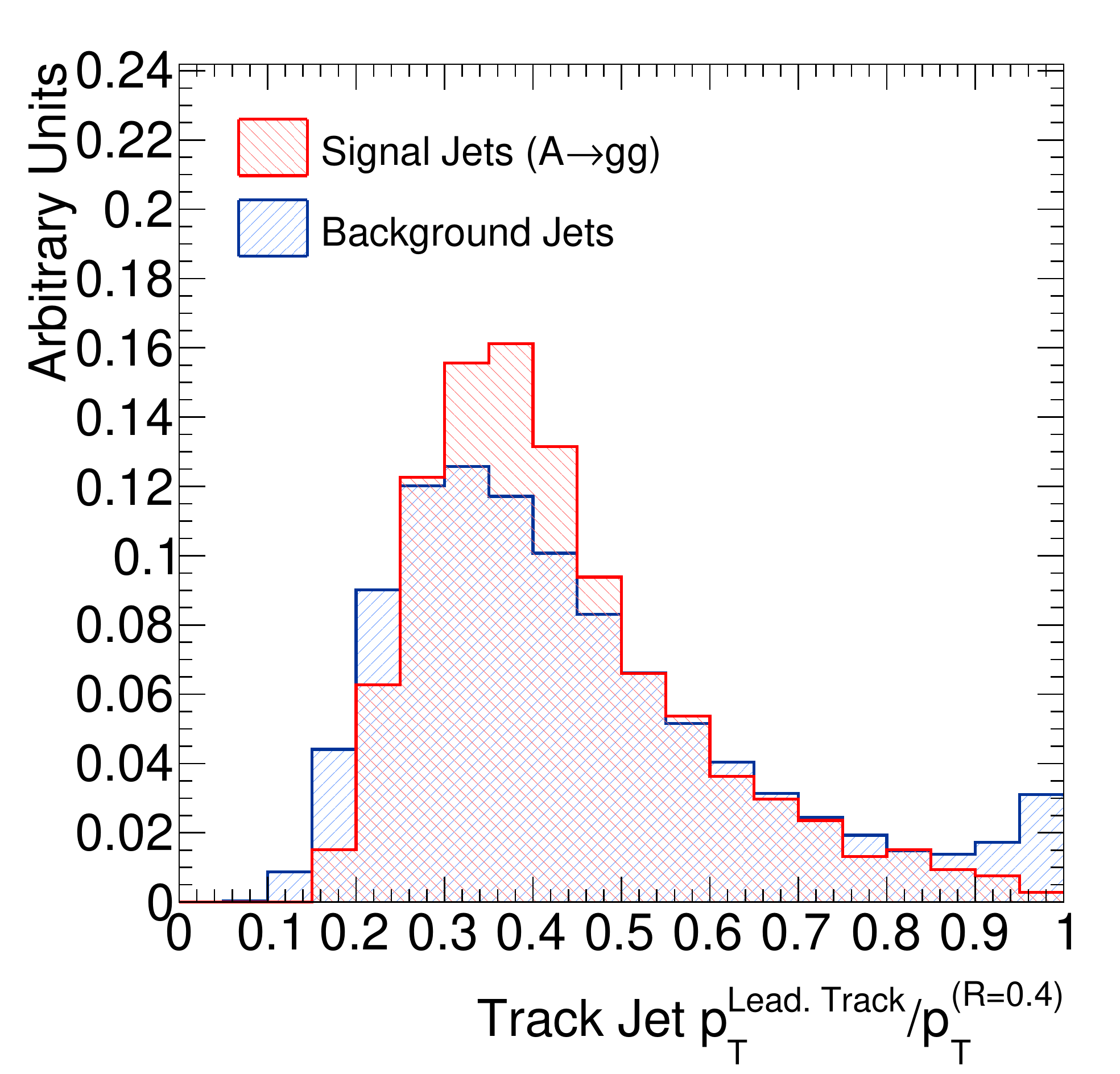}
\caption{Distributions of eight of the variables used as input to the BDT training for jets from
$A\to gg$ decays and jets produced in association with $Z$ bosons.\label{fig:BDTvars}}
\end{figure*}

The performance of the BDT is summarised in Fig.~\ref{fig:roc},
where the background rejection is shown as a function of the signal
efficiency. Higgs decays into a composite light resonance $\eta_c$ and
Higgs decays into an elementary pseudoscalar $A$, which in turn decays
hadronically, are considered separately. For the elementary
pseudoscalar, individual curves for the case in which it decays into a
pair of quarks ($c\bar c$ taken as an example) and for the case in
which in decays into a pair of gluons are shown. These pseudoscalar
decay modes will be of relevance for the interpretation of our results
in the context of 2HDMs in Sect.~\ref{sec:int}. Examples of the
distributions of the variables used to train the BDT are shown in
Fig.~\ref{fig:BDTvars}. The most important variables in terms of
discrimination between signal and background are found to be the jet
masses, followed by the number of track constituents associated with
the track jets.

\subsection{Selection of $h\to ZA$ and $h\to Z\,\etac$ decays}

\begin{figure*}[h]
\centering

\includegraphics[width=0.48\textwidth]{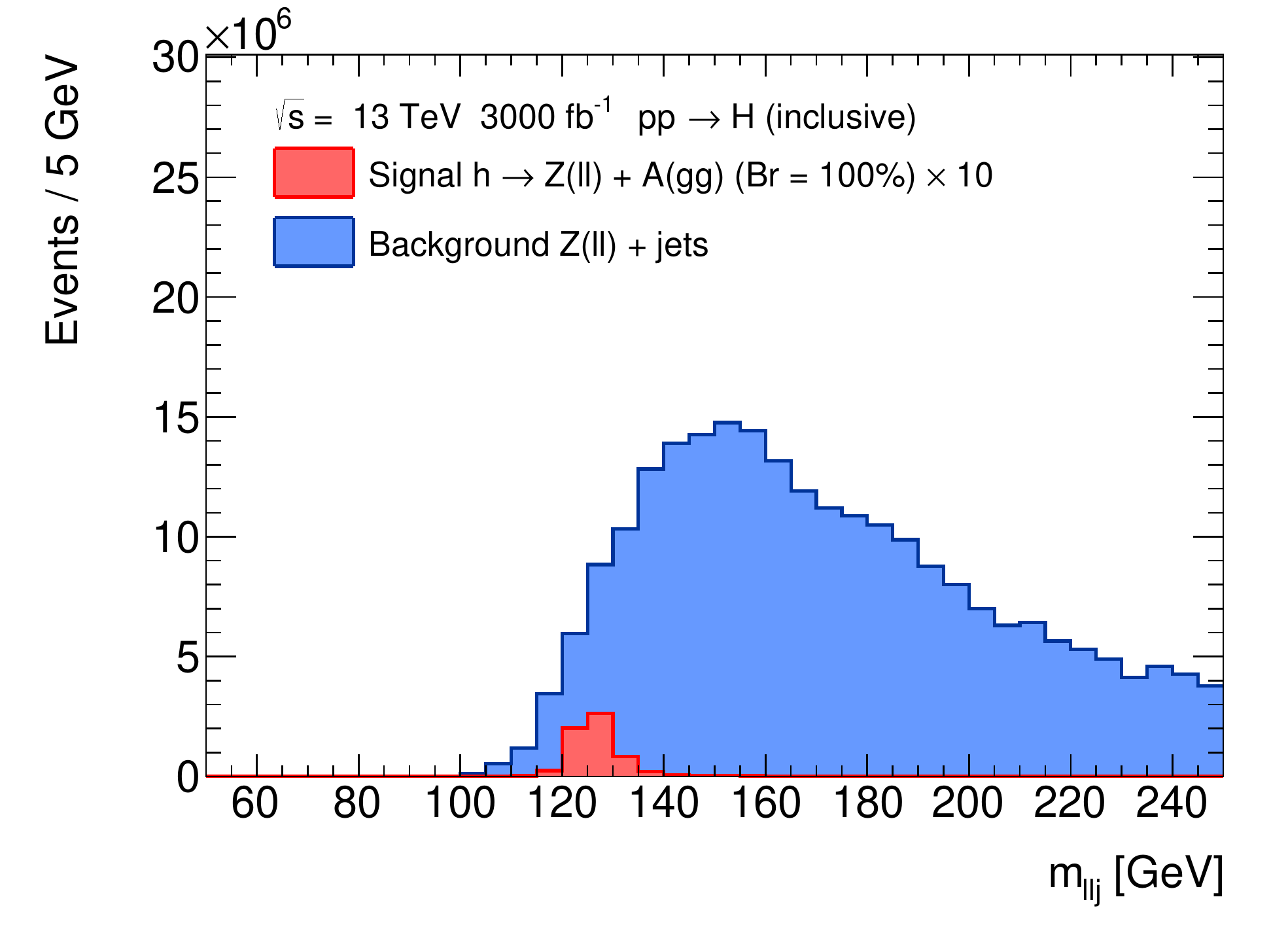}
\includegraphics[width=0.48\textwidth]{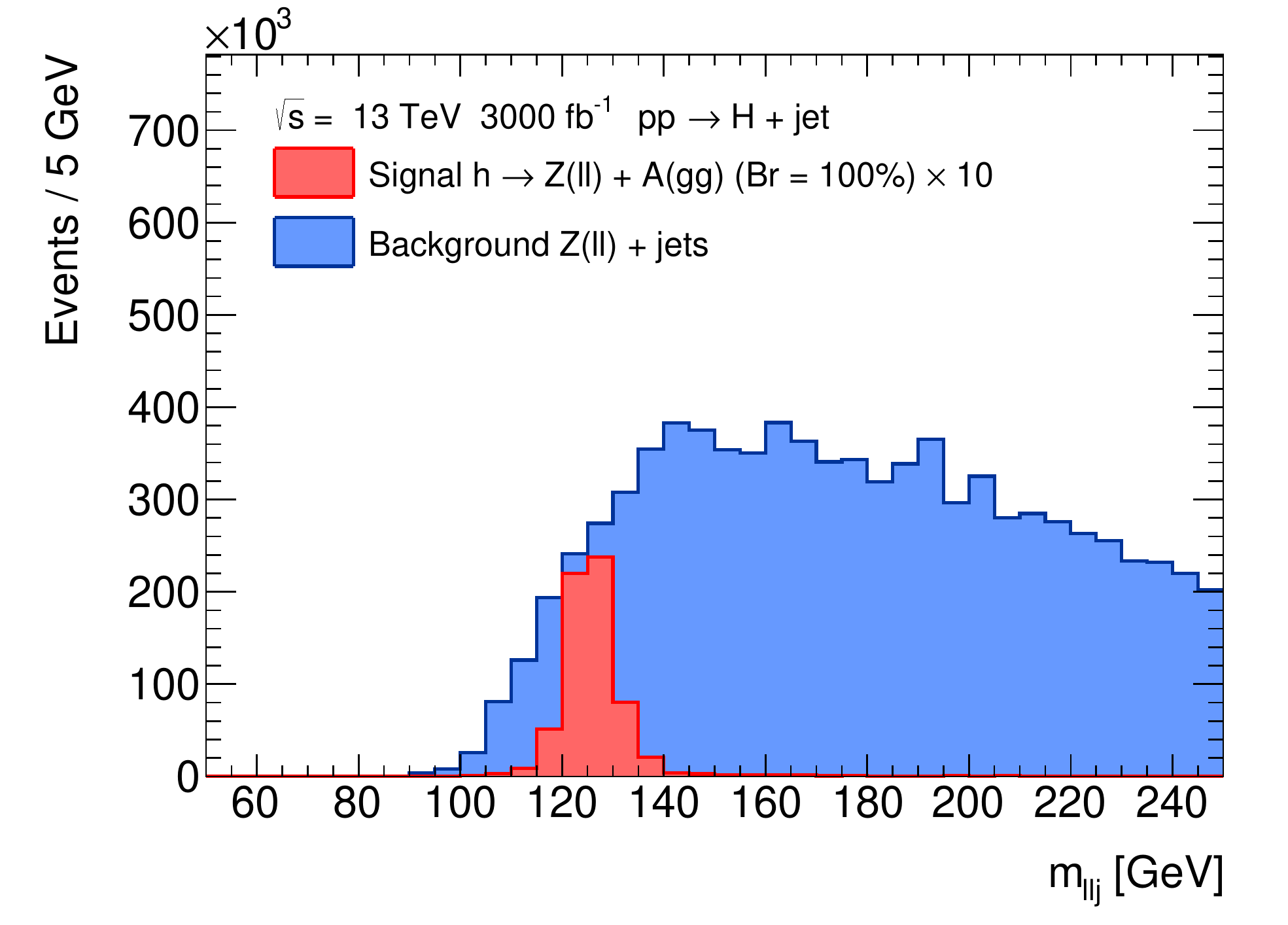}

\includegraphics[width=0.48\textwidth]{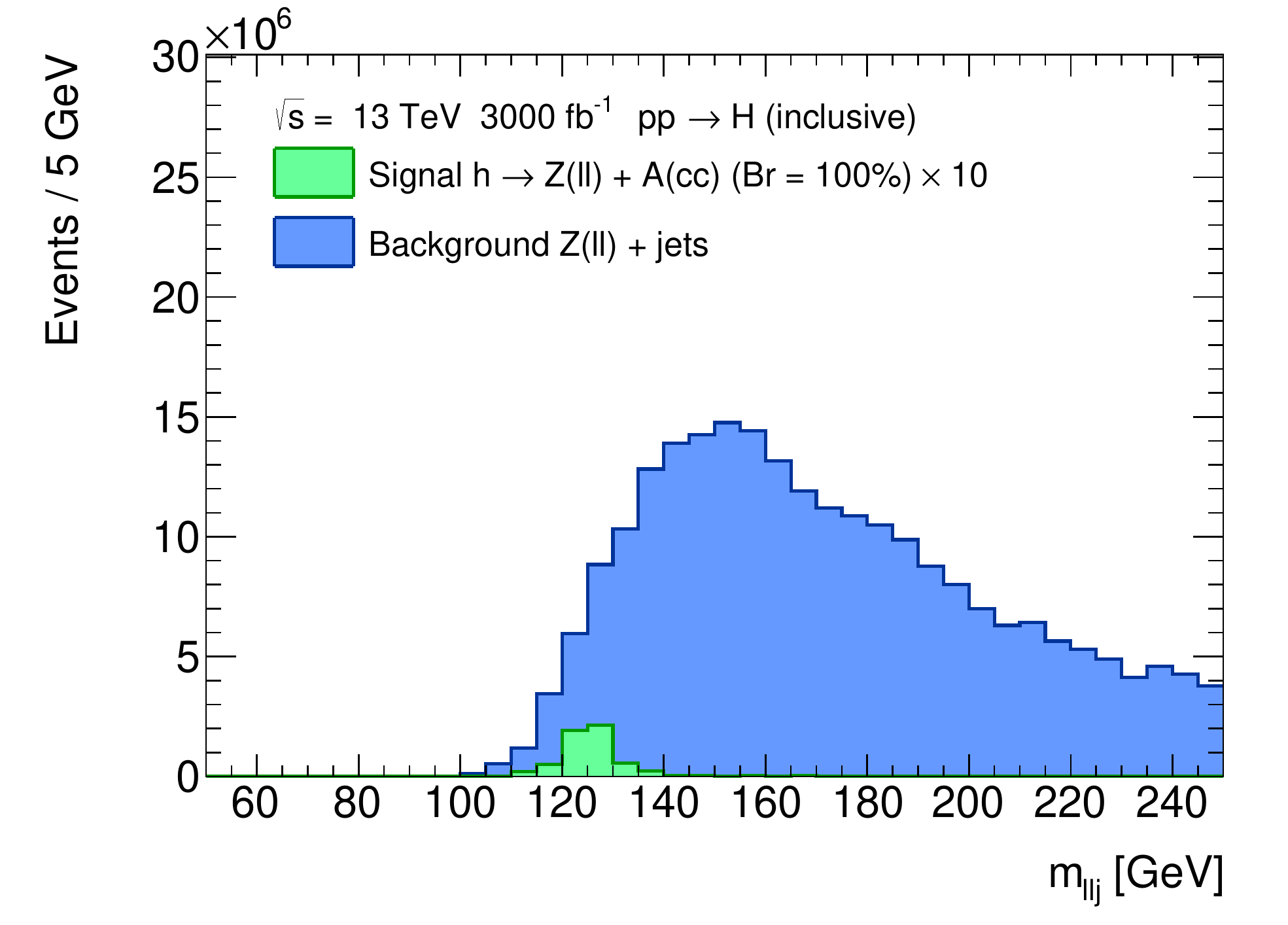}
\includegraphics[width=0.48\textwidth]{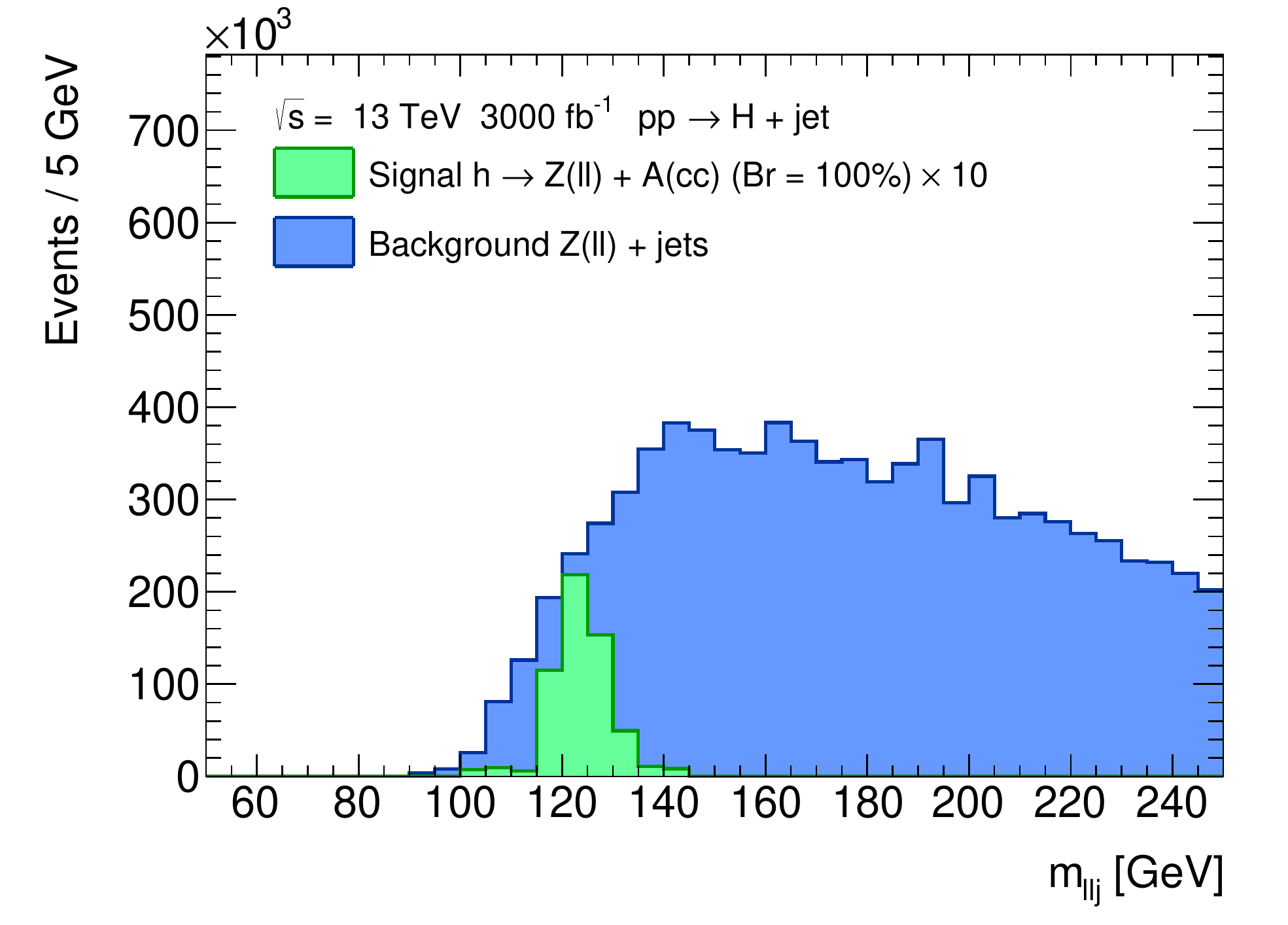}

\includegraphics[width=0.48\textwidth]{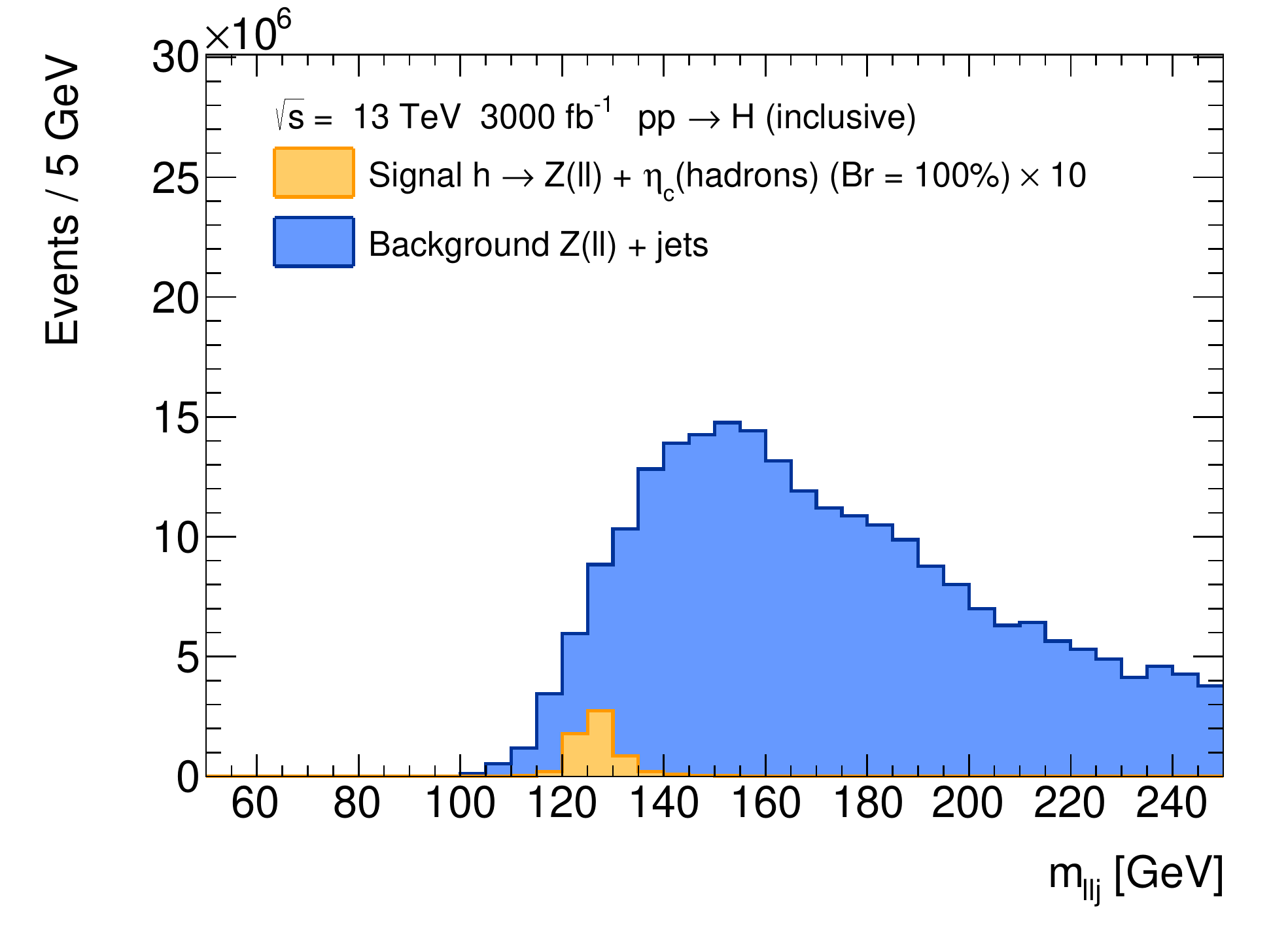}
\includegraphics[width=0.48\textwidth]{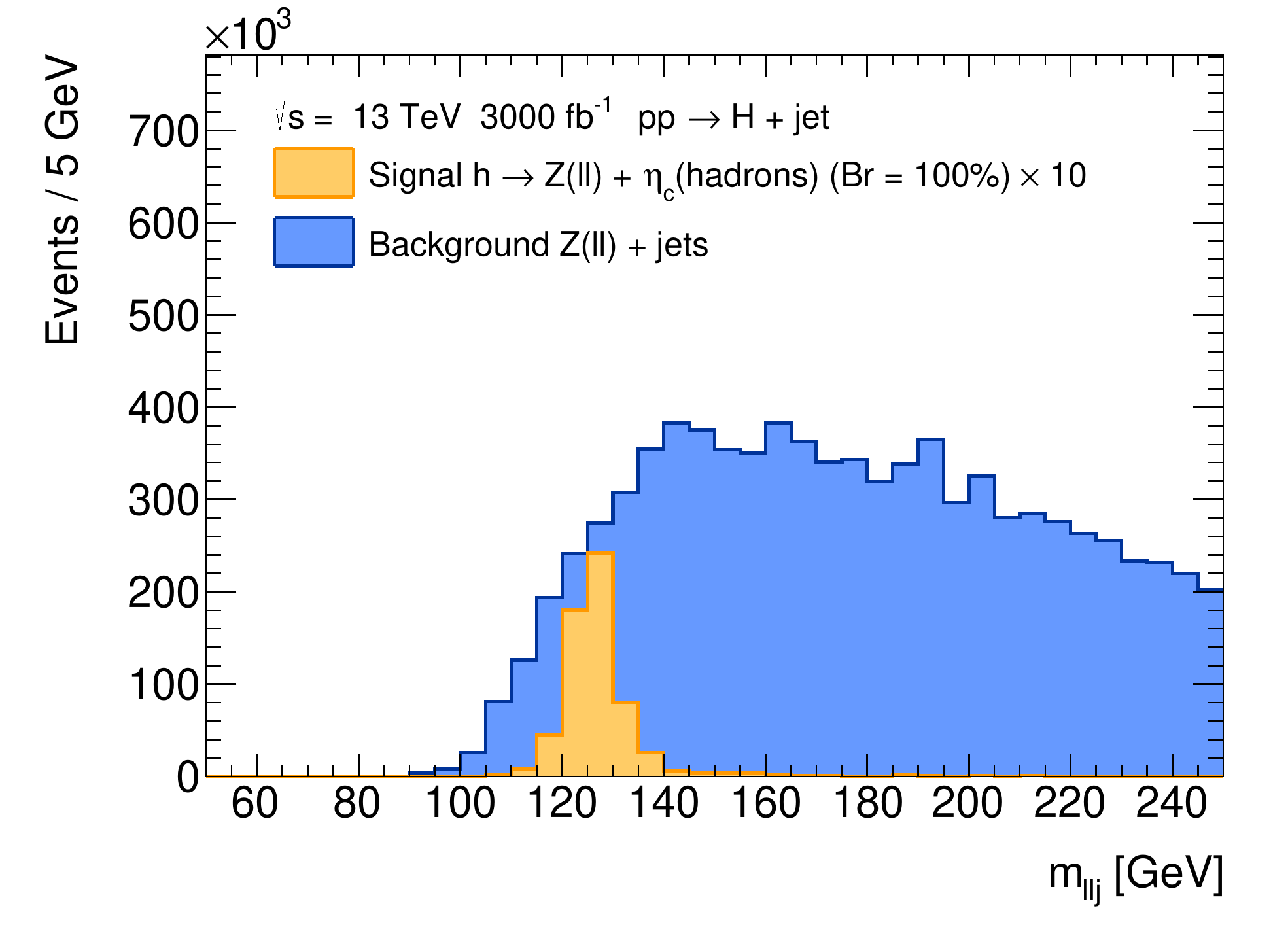}

\caption{The invariant mass distribution of the jet-dilepton system
  (with no BDT based selection applied) in inclusive $h$ production
  (left) and $h+\rm{jet}$ production (right) is shown for $A\to gg$
  (top), $A\to c\bar{c}$ (middle) and $\etac\to\rm{hadrons}$ (bottom)
  signals in comparison to the background
  contribution. The signal contribution is multiplied by ten to improve visibility.\label{fig:mass}}
\end{figure*}

Events containing at least one hadronic decay candidate and one $Z\to\ell\ell$ candidate are considered for further analysis.
In the case of the $h+\rm{jet}$ production channel, an additional $R=0.4$  anti-$k_{t}$ calorimeter jet with $p_{T} > \SI{150}{\giga\electronvolt}$ and $|\eta|<2.5$ is required (no substructure or matching track jet is required).
The single $Z$ boson candidate with $m_{\ell\ell}$ closest to the $Z$ boson mass is chosen to form the $h\to ZA(\etac)$ candidate. If multiple hadronic
decay candidates are reconstructed, the candidate which when paired with the $Z\to\ell\ell$ candidate has an invariant mass closest to $m_{h}=\SI{125}{\giga\electronvolt}$ is chosen.
Finally, the transverse momentum of the $h$ candidate is required to exceed $\SI{20}{\giga\electronvolt}$. The invariant mass of the jet--dilepton system is shown for the inclusive and $h+\rm{jet}$ production channels in Fig.~\ref{fig:mass}.

\begin{figure*}[h]
\centering

\includegraphics[width=0.48\textwidth]{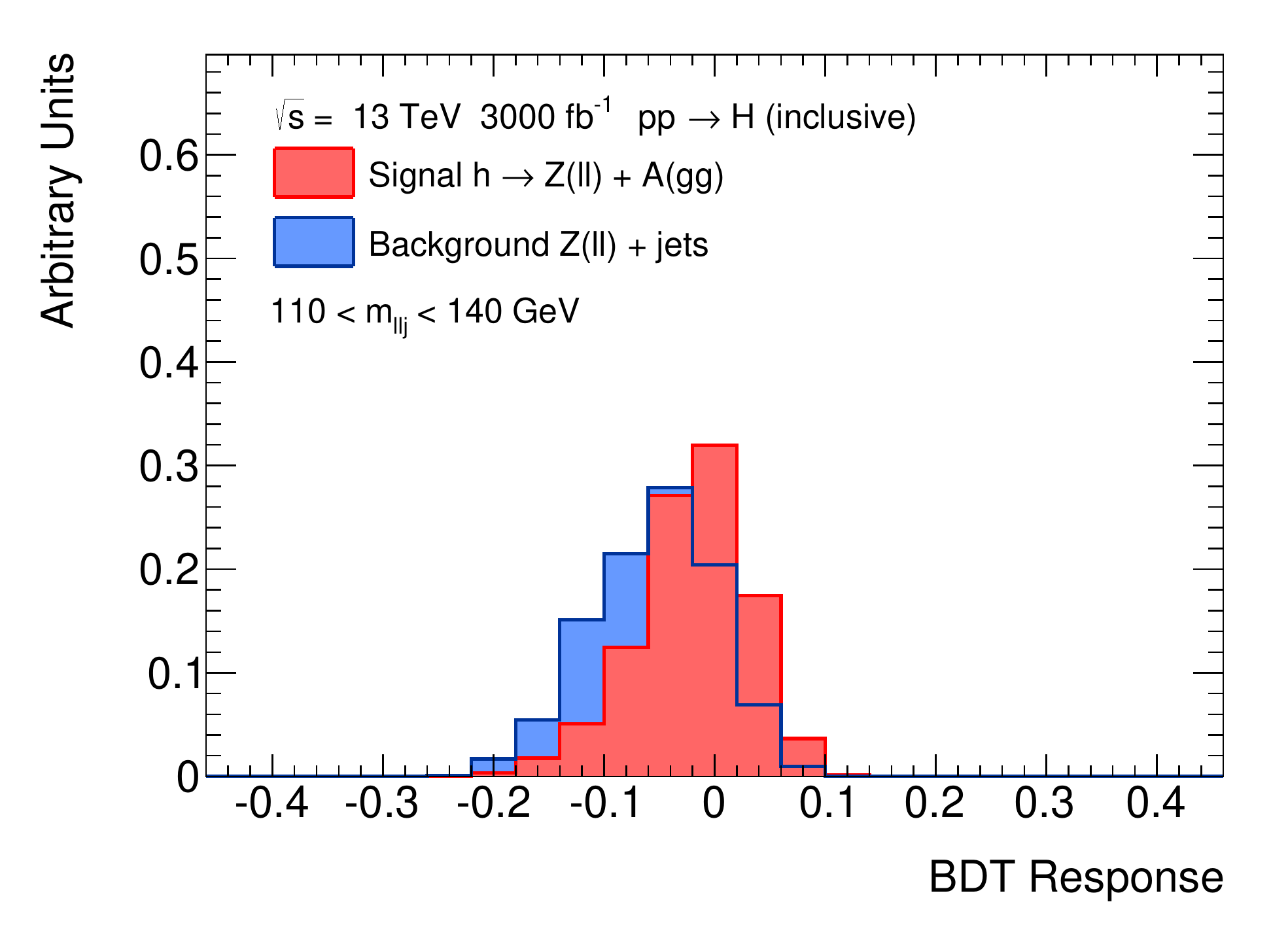}
\includegraphics[width=0.48\textwidth]{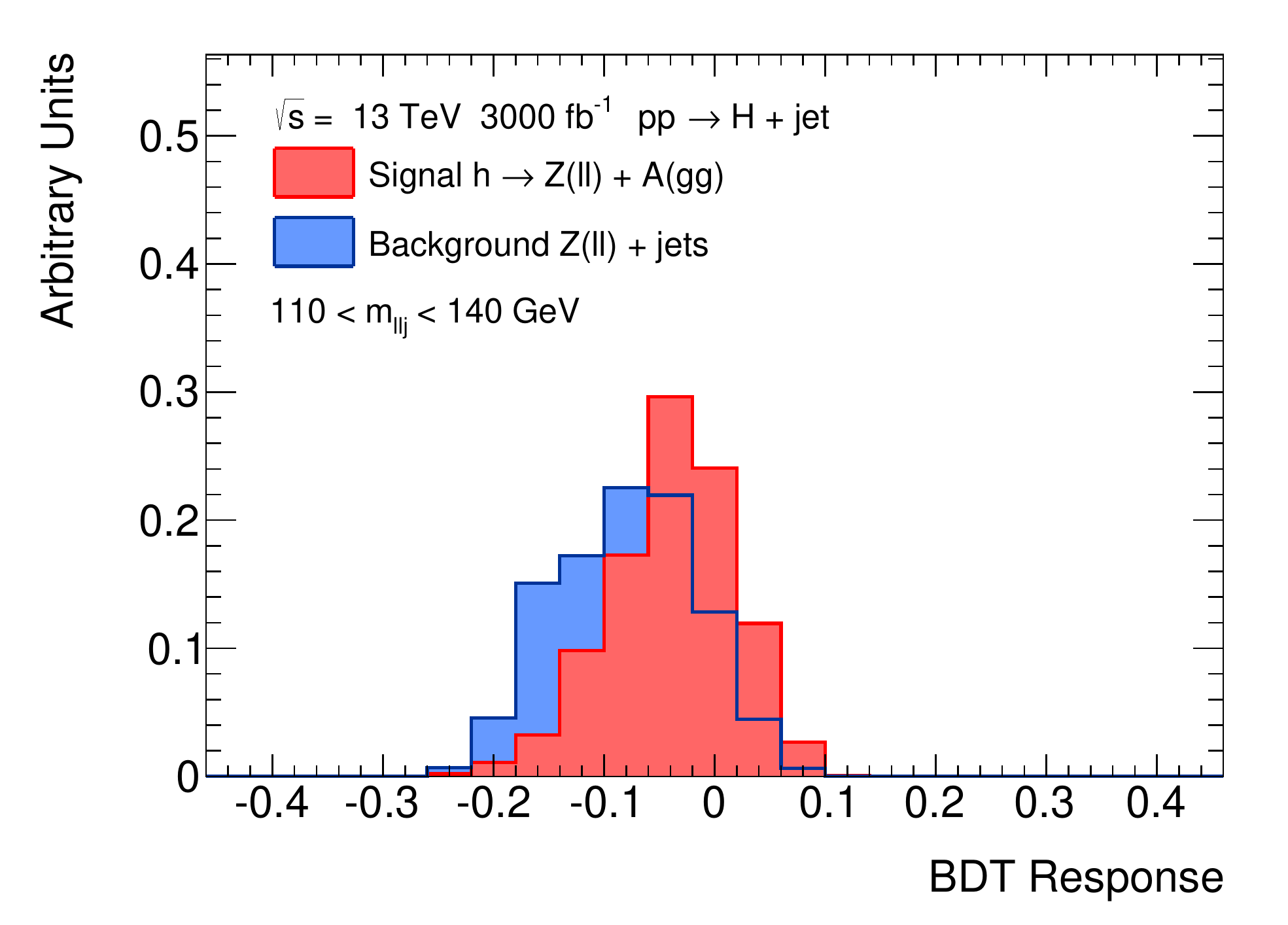}

\includegraphics[width=0.48\textwidth]{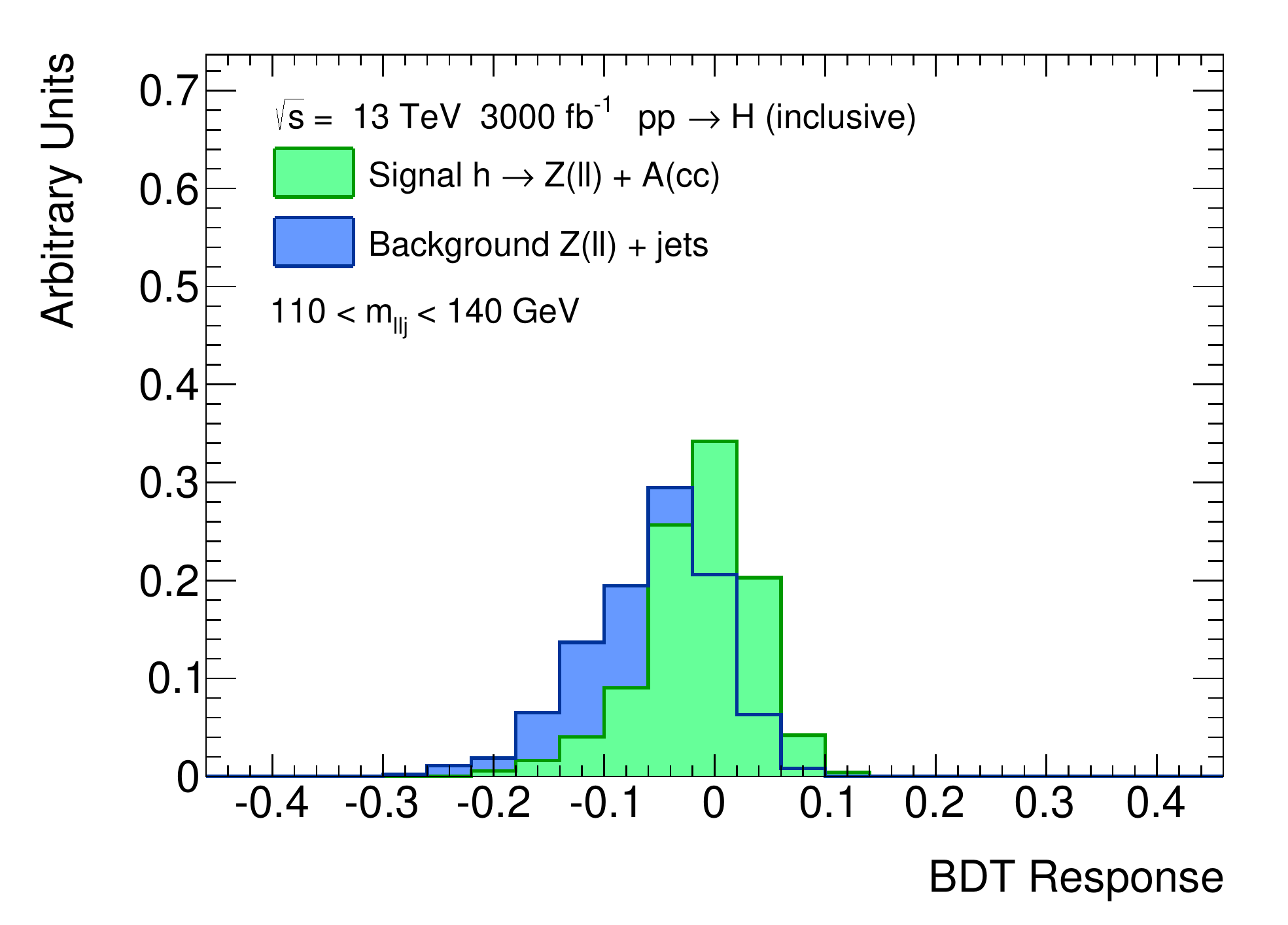}
\includegraphics[width=0.48\textwidth]{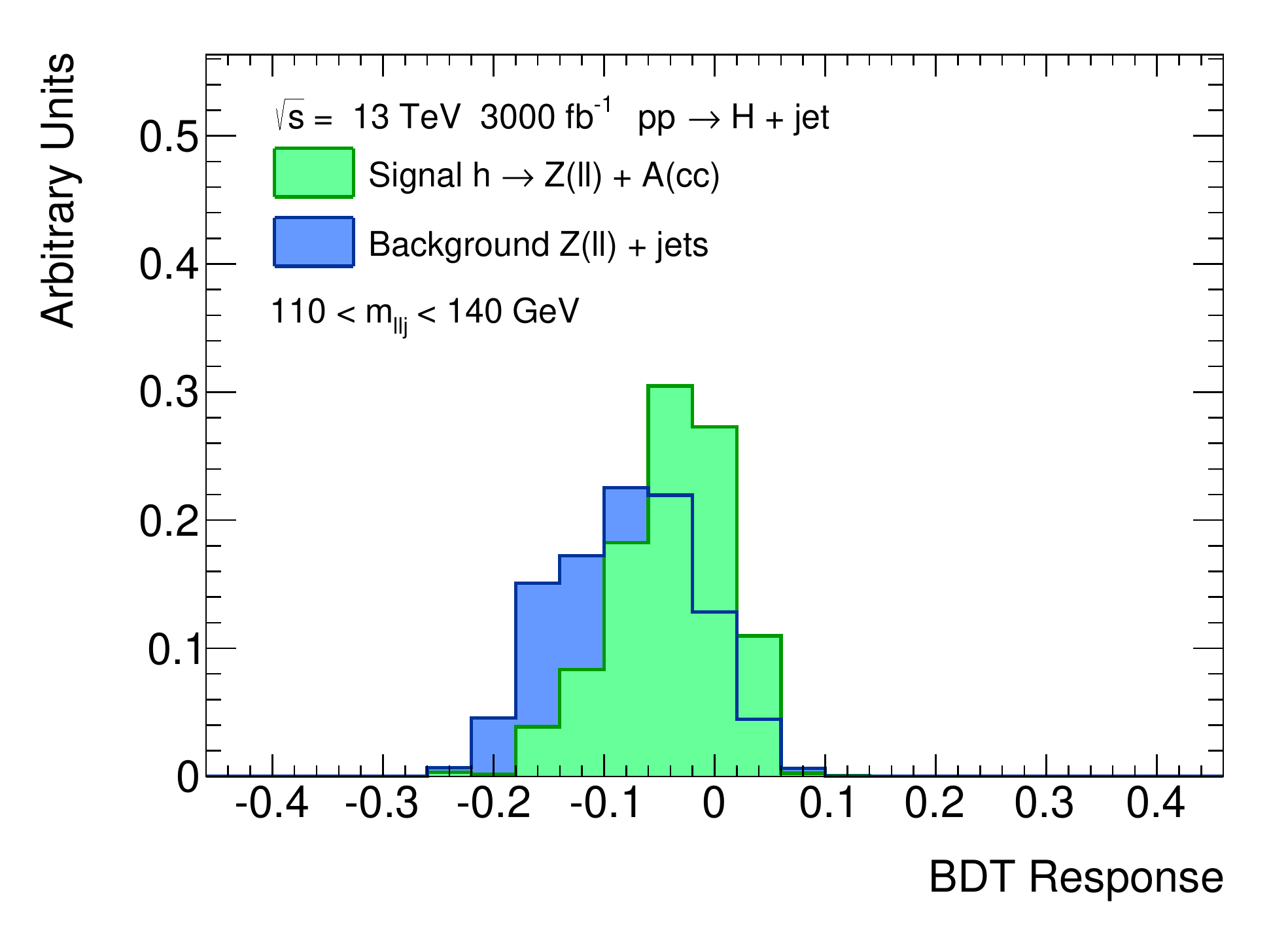}

\includegraphics[width=0.48\textwidth]{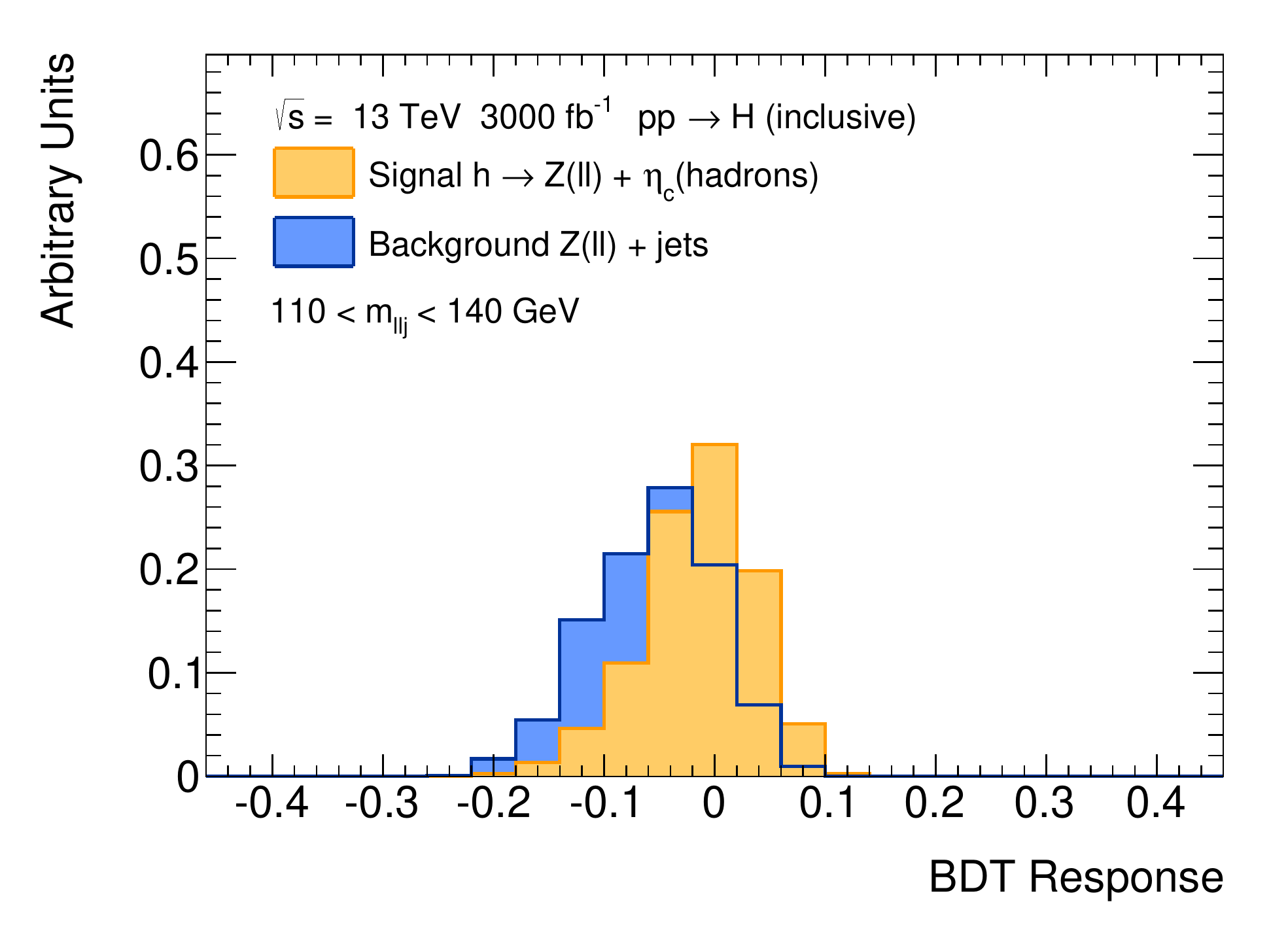}
\includegraphics[width=0.48\textwidth]{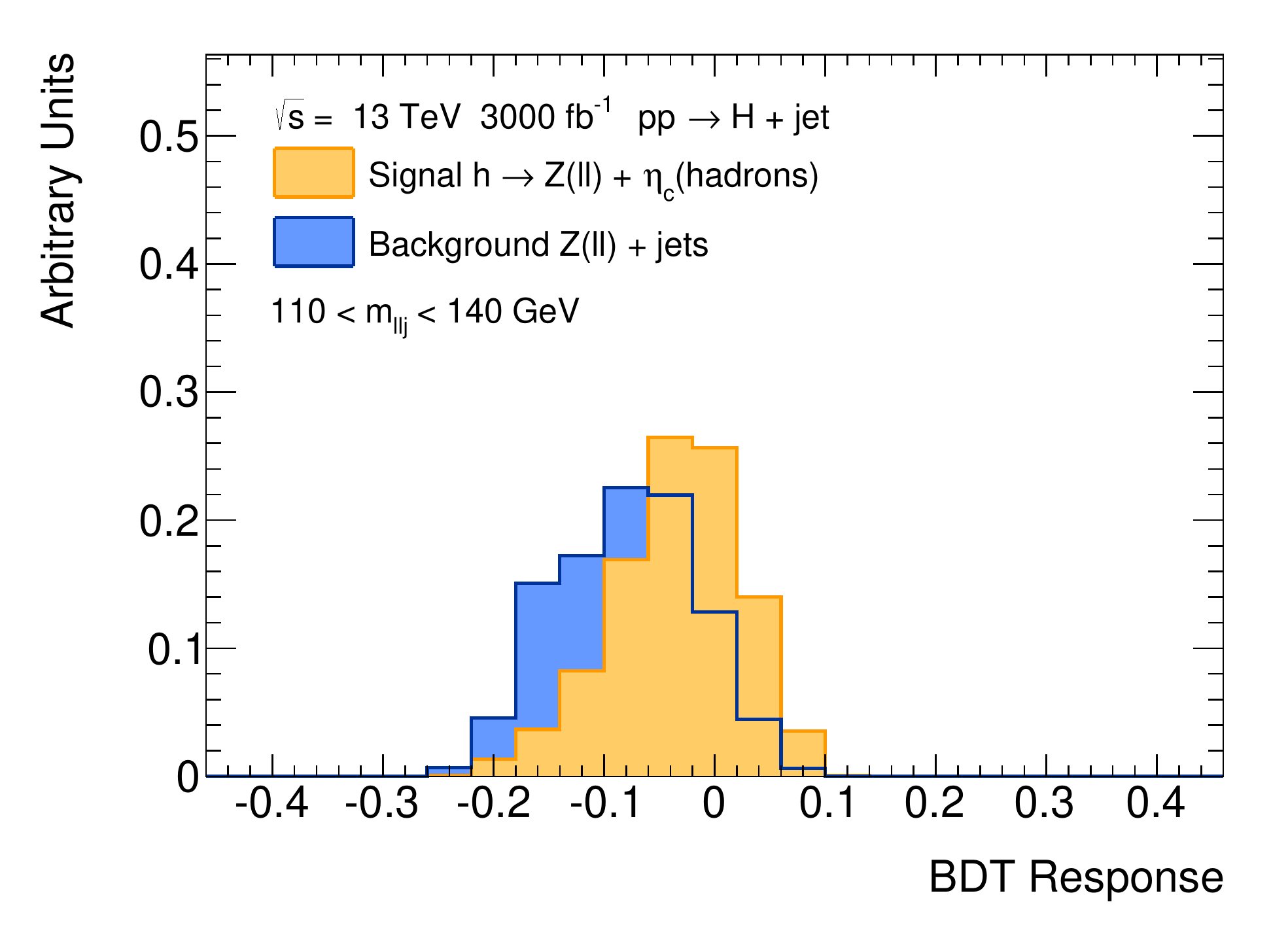}

\caption{The normalised BDT response for the inclusive (left) and $h+\rm{jet}$ (right) production shown
for $A\to gg$ (top), $A\to c\bar{c}$ (middle) and $\etac\to\rm{hadrons}$ (right) in comparison to the background.\label{fig:bdt}}
\end{figure*}

The BDT response is shown for both the signal and the background contributions to the inclusive and $h+\rm{jet}$ production channels in Fig.~\ref{fig:bdt}.

\section{Statistical analysis and results}
\label{sec:results}

The expected performance of the analysis is used to evaluate expected 95~\% CL limits on the branching fractions
 ${\cal BR}\left(h\to ZA\right)$, in the cases where ${\cal BR}\left(A\to gg\right) = 1.0$ or ${\cal BR}\left(A\to c\bar{c}\right) = 1.0$, and ${\cal BR}\left(h\to Z\etac\right)$.
The yields of signal and background events within $\SI{110}{\giga\electronvolt} < m_{\ell\ell j} < \SI{140}{\giga\electronvolt}$ are used to evaluate the limits. To exploit
the additional sensitivity offered by the BDT, a requirement on the BDT response is imposed. The value of this requirement is optimised
to provide the best limit on the branching fractions of interest. The expected 95~\% CL limits on the branching fractions
of interest are shown Table~\ref{tab:limits}. Branching fraction limits at the $1~\%$ level can be expected. The inclusive production channel
is found to be slightly more sensitive than the $h+\rm{jet}$ channel.

In addition to the channels described, Higgs boson production in association with a leptonically decaying $Z$ boson was also considered as a possible
channel to gain additional sensitivity. Initial studies into this channel demonstrated improved signal-to-background ratios when compared to the two channels
constituting the main study, though the substantially lower number of signal events resulted in expected branching fraction limits that were up to an order of magnitude higher than the inclusive and $h+\rm{jet}$ channels.

\begin{table}[h]
\centering
\begin{tabular}{|c|c|c|c|}
\hline
\multirow{2}{*}{Channel}& \multicolumn{3}{c|}{${\cal BR}$ 95~\% confidence level upper limit}\\
\cline{2-4}
& $h\to ZA(\to gg)$ & $h\to ZA(\to c\bar{c})$ & $h\to Z\,\etac$ \\
\hline
Inclusive & $2.0\%$ & $2.1\%$ & $2.0\%$ \\
$h+\rm{jet}$ & $3.5\%$ & $3.9\%$ & $3.7\%$ \\
\hline
\end{tabular}

\caption{The expected 95~\% CL limits on the branching fractions of interest for both the inclusive and $h+\rm{jet}$ channels, assuming \SI{3000}{\per\fb}
at $\sqrt{s}=\SI{13}{\tera\electronvolt}$.\label{tab:limits}}
\end{table}

\section{Constraints on the 2HDM parameter space}
\label{sec:int}

With a focus on the HL-LHC, we assume the Higgs boson couplings to be
tightly constrained to SM-like values. Assuming no evidence for new
physics in the HL-LHC data, any 2HDM scenario compatible with the
observations would therefore necessarily be close to the alignment
limit. It has been pointed out in Ref.~\cite{Bernon:2015qea} that a
light pseudoscalar $A$ with mass below \SI{10}{\giga\electronvolt} can
be accommodated in this limit, particularly in Type I models, which we
consider here. A pseudoscalar that light can decay into pairs of
fermions through tree-level interactions or into pairs of gluons and
photons through loop-induced couplings. In Type I models, the
tree-level couplings to fermions are essentially given by the fermion
masses times a universal factor of $\cot(\beta)$. A considerable
hadronic branching fraction hence arises from decays into quark pairs,
gluon pairs, or indirectly from decays into pairs of tau leptons that
decay into hadrons subsequently. As shown in Fig.~\ref{fig:roc}, the
performance of our analysis is fairly insensitive to the details of
the hadronic decay mode of the pseudoscalar. The results of our
analysis can therefore directly be used in order to constrain such
models. To the best of our knowledge, no detailed analysis of this
final state has been provided in the literature so far.

In order to assess the constraining power of our results, we perform a
parameter scan for a fixed benchmark pseudoscalar mass of
$m_A{=}\SI{4}{\giga\electronvolt}$. For the chosen benchmark value of
$m_A$, decays into tau leptons and charm quarks dominate. Decays into
gluon pairs contribute a branching fraction at the per cent level.
Overall, we obtain
$\mathcal{BR}(A\to \mathrm{hadrons}){\approx}\SI{82}{\percent}$.

In our parameter scan, we calculate the branching ratio relevant for
the interpretation of our results, $\mathcal{BR}(h\to ZA)$, for each
parameter point. The corresponding partial decay width is given by
\begin{align}
  \Gamma(h\to ZA) &= \frac{|\vec{p}|}{8\pi m_h^2}|\mathcal{M}(h\to
  ZA)|^2 = \frac{g_{hZA}^2}{2\pi}\frac{|\vec{p}|^3}{m_Z^2},
  \label{eq:gam_hza}
\end{align}
at tree level, where $\vec{p}$ is the three-momentum of either of the
two decay products in the rest frame of the Higgs boson. The
$hZA$-coupling is given by
\begin{align}
  g_{hZA} = \frac{e\cos(\beta-\alpha)}{2\cos\theta_W\sin\theta_W}.
\end{align}
The partial decay width $\Gamma(h\to ZA)$ therefore vanishes in the
strict alignment limit with $\cos(\beta-\alpha)=0$. The corresponding
branching fraction, however, becomes sizable already for small
$\cos(\beta-\alpha)$ if the decay $h\to AA$ does not contribute substantially
to the Higgs boson total width. We therefore focus on the
parameter region, where $g_{h_\mathrm{SM}AA}=0$ at tree level, which
implies~\cite{Casolino:2015cza}
\begin{align}
  m_{12}^2 = (2m^2 + m_h^2)\sin(2\beta)/4.0.
\end{align}
To ensure alignment, we perform a uniform scan with
$\sin(\beta-\alpha)\in[0.99,1.0]$. In this regime, we can assume the
production cross sections of the \SI{125}{\giga\electronvolt} Higgs to
be SM-like and directly apply our previously obtained limit on
$\mathcal{BR}(h\to ZA)$. Note, however, that the limit must be applied
to $\mathcal{BR}(h\to ZA)\times\mathcal{BR}(A\to \mathrm{hadrons})$,
since $\mathcal{BR}(A\to \mathrm{hadrons})=1$ was assumed previously.
The remaining free parameters of the model are uniformly varied in the
intervals $m_H\in[130,600]\si{\giga\electronvolt}$,
$m_{H^\pm}\in[50,600]\si{\giga\electronvolt}$, and
$\tan\beta\in[0.1,5.0]$. We calculate the physical spectrum and the
relevant branching fractions with 2HDMC version
1.7.0~\cite{Eriksson:2009ws}.

For each point we check for vacuum stability of the potential,
tree-level unitarity using the corresponding functionalities of 2HDMC.
On the phenomenological side, we check for compatibility of the
resulting oblique parameters $S,T,U$~\cite{Peskin:1990zt,Peskin:1991sw}, as calculated by 2HDMC, with
electroweak constraints~\cite{Baak:2014ora}. Only points that can be
accommodated within these constraints are retained. Points that are
incompatible with exclusion limits set by LEP, Tevatron, and LHC
analyses are also rejected. For this purpose, we employ numerical
program HiggsBounds~\cite{Bechtle:2008jh,Bechtle:2011sb,Bechtle:2013gu,Bechtle:2015pma}
and include all analyses implemented in version 4.3.1. Only parameter
points for which none of the scalars in the spectrum can be excluded
at \SI{95}{\percent} CL are retained in our scan. In order to check
the compatibility with the LHC and Tevatron Higgs boson signals in our
scan, we employ the HiggsSignals program
\cite{Bechtle:2013xfa,Bechtle:2014ewa} version 1.4.0. We discard any points that are
excluded at \SI{95}{\percent} confidence level based on the $\chi^2$
calculated by HiggsSignals.

\begin{figure}[h!]
  \centering
  \includegraphics[width=.45\textwidth]{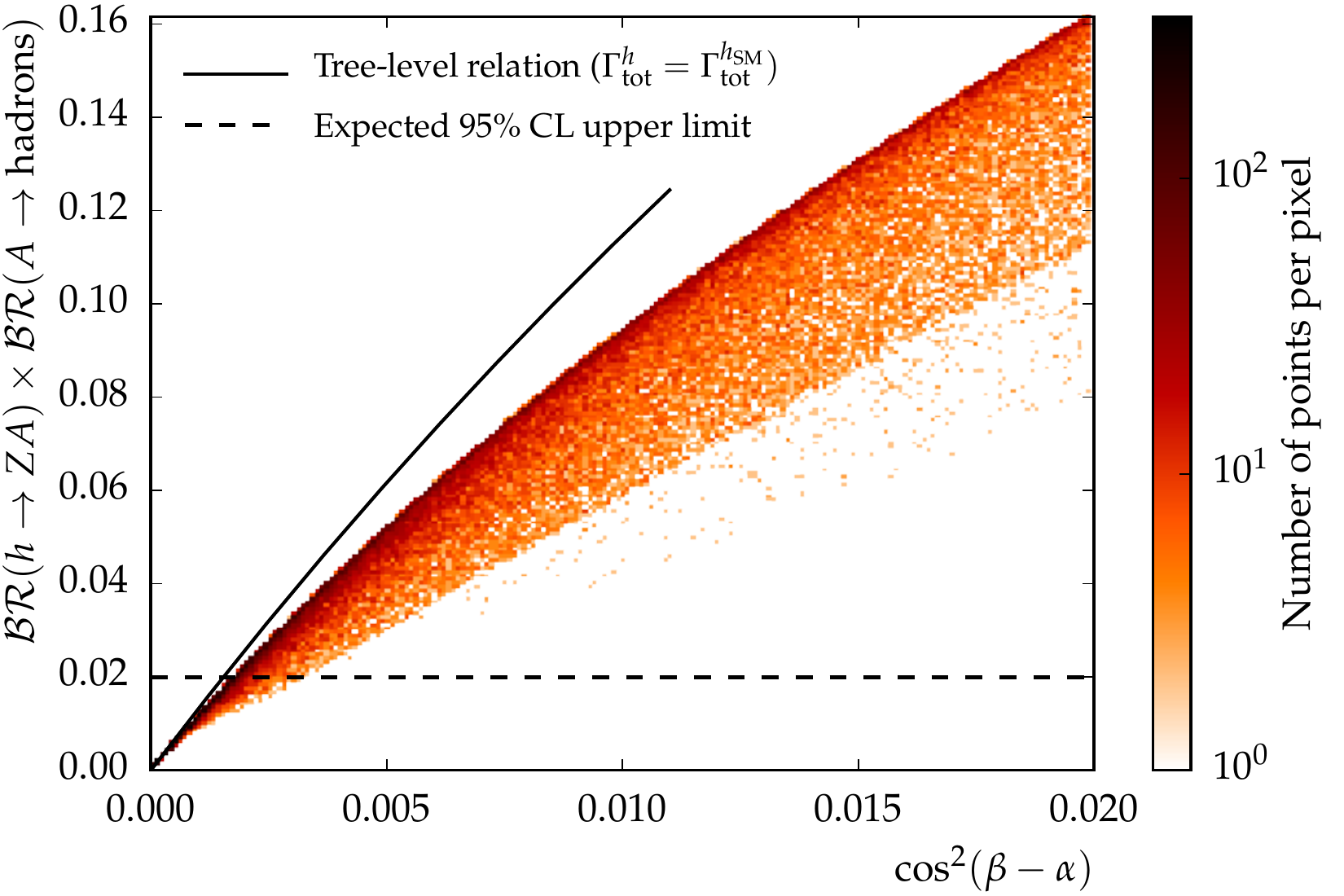}
  \caption{Distribution of scanned parameter points in the
    $\cos^2(\beta-\alpha)$ vs. $\mathcal{BR}(h\to
    ZA)\times\mathcal{BR}(A\to \mathrm{hadrons})$ plane. The
    color-coding denotes the density of points in the respective areas
    as indicated by the color bar. We also display the tree-level
    functional relationship between $\cos^2(\beta-\alpha)$ and
    $\mathcal{BR}(h\to ZA)\times\mathcal{BR}(A\to \mathrm{hadrons})$,
    assuming
    $\Gamma^h_\mathrm{tot}=\Gamma^{h_\mathrm{SM}}_\mathrm{tot}$. The
    dashed line shows the expected \SI{95}{\percent} CL upper limit on
    the displayed branching fraction. All points above this line are
    expected to be excluded by the analysis presented here.}
  \label{fig:2hdm_cba2}
\end{figure}

In Fig.~\ref{fig:2hdm_cba2}, we illustrate the results of the
parameter scan. We display the distribution of all parameter points
that pass the applied theoretical and phenomenological constraints in
a two-dimensional parameter plane spanned by $\cos^2(\beta-\alpha)$
and $\mathcal{BR}(h\to ZA)\times\mathcal{BR}(A\to \mathrm{hadrons})$
along with the tree-level functional dependence of these quantities
given by~Eq.~\eqref{eq:gam_hza}, assuming for simplicity
$\Gamma^h_\mathrm{tot}=\Gamma^{h_\mathrm{SM}}_\mathrm{tot}$. For large
$\cos(\beta-\alpha)$, this is assumption is violated due to the
opening of further decay channels. At small $\cos(\beta-\alpha)$,
however, the corresponding approximation proves to be reasonable for
parameter points that pass the applied phenomenological constraints.
As illustrated in Fig.~\ref{fig:2hdm_cba2}, the scanned parameter
space can effectively be constrained to very small values of
$\cos^2(\beta-\alpha)$ by applying our expected limit on
$\mathcal{BR}(h\to ZA)$. In fact, we find that no parameter point with
$\cos^2(\beta-\alpha)>\num{0.0035}$ survives the limit set by the
analysis presented above, translating to $\sin(\beta-\alpha)\gtrsim
0.998$ in the scanned subspace of parameters. Correspondingly, a mere
\SI{12}{\percent} of the parameter points displayed in Fig.~\ref{fig:2hdm_cba2} fall in the region of allowed values for
$\mathcal{BR}(h\to ZA)\times\mathcal{BR}(A\to \mathrm{hadrons})$ after
applying the limit on $\mathcal{BR}(h\to ZA)$ obtained above.

\section{Summary}
\label{sec:sum}

Searches for rare and exclusive Higgs boson decays are at the core of the program of the High Luminosity LHC. The observation of Higgs boson decays into light elementary or composite resonances would be evidence for the existence of physics beyond the Standard Model.

While previous experimental strategies to reconstruct light resonances relied entirely on their leptonic decay products, in this work, we evaluated the prospects for their discovery in the often dominant hadronic decay channels. We have focused on the Higgs boson production processes with the largest cross sections, $pp \to h$ and $pp \to h+\mathrm{jet}$, with subsequent decays $h \to ZA$ or $h \to Z\,\etac$. The former is present in many multi-Higgs extensions of the Standard Model, while observing the latter at a branching ratio of $\mathcal{BR}(h \to Z\,\etac) \geq 10^{-3}$ could indicate an enhanced Higgs-charm coupling.

The decay products of light resonances with masses below a few \si{\giga\electronvolt} that arise from Higgs decays are highly collimated, i.e. they get emitted into a small area of the detector. In such scenarios jet substructure is an indispensable tool to retain sensitivity in discriminating signal from large QCD-induced backgrounds. In particular, by exploiting the improved angular resolution of track-based observables, a good signal-to-background discrimination can be achieved, which results in a limit on the branching ratios of $\mathcal{O}(1)~\%$ for a data sample corresponding to $\SI{3000}{\per\fb}$ at $\sqrt{s}=\SI{13}{\tera\electronvolt}$.

\section*{Acknowledgments}

SK's work was supported by the European Union as part of the FP7 Marie
Curie Initial Training Network MCnetITN (PITN-GA-2012-315877). AC and
KN are supported in part by the European Union's FP7 Marie Curie
Career Integration Grant EWSB (PCIG12-GA-2012-334034). This work was
made possible thanks to the Institute of Particle Physics Phenomenology
Associate scheme.

\bibliographystyle{JHEP}
\bibliography{paper}

\end{document}